\documentclass[astrosymb, tighten, preprint, 3p]{cas-dc}

\usepackage[]{natbib}
\usepackage{graphicx}
\usepackage{amsmath}
\usepackage{pifont}
\usepackage[section]{placeins}

\usepackage{subfig}

\ExplSyntaxOn
\cs_gset:Npn \__first_footerline:
  { \group_begin: \small 
  Author\, Preprint\, from\, {\it Icarus} (2007)
  \group_end: }
\ExplSyntaxOff 

\makeatletter
\newcommand\notsotiny{\@setfontsize\notsotiny\@vipt\@viipt}
\makeatother

\begin{document}

\title{\Huge A survey of debris trails from short-period comets}

\author[1]{William T. Reach}
\affiliation[1]{California Institute of Technology, Pasadena, CA, USA}
\ead{wreach@spacescience.org}

\author[2]{Michael S. Kelley}
\affiliation[2]{University of Minnesota, Minneapolis, MN, USA}

\author[3]{Mark V. Sykes}
\affiliation[3]{Planetary Science Institute, Tucson, AZ, USA}

\newcommand{\ISO}{{\it ISO}}
\newcommand{\etal}{ {\it et al.} }
\newcommand{\simgt}{\ga}
\newcommand{\simlt}{\lower.5ex\hbox{$\; \buildrel < \over \sim \;$}}

\def\nb{\noindent{\bf N.B.}}

\shorttitle{Cometary Debris Trails}
\shortauthors{Reach {\it et al.}}

\def\nIRtrail{30}   
\def\nSpitDicov{23} 

\def\ncomobs{34} 
\def\ncomtrail{27} 
\def\nconfuse{4} 
\def\nNOtrail{3} 
\def\nintermed{4}
\def\pertdateone{35 yr}

\def\ntrail{30}
\def\ntrailnew{22}


\begin{abstract}
We observed  \ncomobs\ comets using the 24 $\mu$m
camera on the {\it Spitzer Space Telescope}. Each image contains the nucleus
and covers at least $10^6$ km of each comet's orbit. Debris trails due to
mm-sized or larger 
particles were found along the
orbits of \ncomtrail\ comets; \nconfuse\ comets had small-particle dust tails
and a viewing geometry that made debris trails impossible to distinguish; 
and only \nNOtrail\ had
no debris trail despite favorable observing conditions. 
There are now \ntrail\ Jupiter-family comets with known debris trails, of 
which \ntrailnew\ are reported in this paper for the first time.
The detection rate is $>80$\%, indicating that debris trails are 
a generic feature of short-period comets.
By comparison to 
orbital calculations for particles of a range of sizes ejected over 2 yr
prior to observation, we find that particles comprising \nintermed\ debris trails
are typically mm-sized  
while the remainder of the debris trails require particles larger
than this. The lower-limit masses of the debris trails are typically $10^{11}$ g,
and the median mass loss rate is 2 kg/s. The mass-loss rate in
trail particles is comparable to that inferred from OH production rates
and larger than that inferred from visible-light scattering in comae.
\end{abstract}

\maketitle

\section{Introduction}

The composition and structure of cometary nuclei remains largely 
unknown. 
Most information about comets derives from 
their two best-determined observable properties: their orbits and the
apparent nature of the material they eject when they approach the Sun.
The nuclei themselves are very dark, offering few clues to their
nature other than their size, shape, and color \citep{LamyNuc,WeisNuc}.
Understanding cometary structure and composition is key to
understanding the formation of the outer planets and the origin
of the outermost layers, especially oceans and organic material,
on Earth \citep{chyba90,cometsoriginlife}.
Specifically, the Earth's outermost layers and oceans are likely to
have been deposited by the late bombardment of the early Earth
by cometary bodies, if we define comets as those bodies on which
water was in icy form at the time of initial planetesimal/cometesimal
coagulation in the solar nebula  \citep{delsemme}.
The cause of the well-known 1908 Tunguska event 
\citep{turco} has been debated as 
an asteroidal or cometary disruption in the atmosphere \citep{chyba93}, 
but how well
do we really know the difference between asteroids and comets?
The cometary bodies left over from the formation of
the cores of Jupiter and Saturn may have been
largely scattered to the Oort cloud, while the remaining comets that formed
Uranus and Neptune, as well as their (and some of Saturn's) numerous moons,
are likely residents of the present Kuiper Belt. 
The Kuiper Belt is the dynamical origin of the present
Jupiter-family, also called `short-period,' comets
\citep{levison}, whose nature we will explore in the present work.

The most significant breakthrough in understanding the physical nature
of comets was the revolutionary work by \citet{whippleI}, in which he demonstrated that
cometary orbital perturbations (beyond those that could be explained
by the major planets) were caused by non-gravitational 
forces due to material ejected by sublimating ices heated by the Sun
during perihelion passages. 
The model has now advanced to include anisotropic
emission from hot-spots or `jets' on rotating nuclei, which
are required both to provide the non-gravitational forces that
change comets' orbits from revolution to revolution as well
as the fan-shaped comae \citep{sekanina79,sekanina}.
Whipple's model of the comet is often described as a `dirty snowball,'
a nickname that suggests several prevailing beliefs about cometary
nature. Foremost, the nickname promotes the critical role played
by icy material in powering the mass-loss that is the defining
characteristic of comets. The nickname also suggests that solid material
is a relatively minor constituent, and that it is in the form of
small particles scattered throughout a loosely agglomerated 
set of ice crystals. This belief, however, is not borne out by
observational studies of comets, including {\it in situ}
studies by spacecraft \citep{keller}, remote infrared observations
\citep[e.g.]{SykesWalker}, and the observations presented in this paper.

Whipple himself frequently mentioned and wrote about the connection
between comets and meteors. In paper II \citep{whippleII}
of his classic series,
he describes how solid particles are ejected
from comets. An example in that paper estimates the ejection
velocity of particles with 1 cm radius as 3 m~s$^{-1}$ and
associates such particles with photographic meteors. 
The mass fraction of rock versus ice in cometary
nuclei was unknown at the time of the development of Whipple's 
comet model. In paper III of his series, he adopted a
20\% fraction of cometary mass in the form of solid material.
However, the model does not require a particularly high or low 
ice content in order to explain the properties of cometary orbits
or mass loss. 

Modern views \citep[e.g.][]{prialnik} of comets recognize the processing and evolution
of cometary material both while in the outer solar system and after
capture into orbits that characterize short-period comets.
The water-powered dirty snowball model could not explain why
many comets are active at great distances \citep{meech}, or why activity is
not restricted to being very close to the Sun for older comets.
With the observation of significant activity in jets, the evolution of comets
and their devolatilization is complicated by the details of composition and the
location of active regions on a comet's surface coupled with the orientation of
the comet spin axis. Comets possessing highly volatile ices such as CO may be 
observed to be active at great heliocentric distances. While some comets can 
decrease in their activity as they approach perihelion as jets 
become shadowed from the Sun.
This makes the detailed characterization of comet activity as a group 
difficult.

The current conceptual model has comets devolatilizing even
while in the outer Solar System, then more rapidly upon
entering a low-perihelion orbit. Sublimating gases carry away
particles small enough that their drag forces exceed gravitational
pull of the nucleus \citep{whippleII}:
\begin{equation}
s < 19 f_A  / r^{9/4} R,
\end{equation}
where $s$ is the radius of the particle (cm), 
$f_A$ is the fraction of the comet surface with exposed ice
(taking the latent heat to be that of H$_2$O), 
$r$ is the distance from the Sun (AU), the density
of the cometary nucleus is taken to be 1 g~cm$^{-3}$,
and $R$ is the nuclear radius (km).
For comets with radii of order 1 km producing dust around 1 AU from the
Sun using ice covering 10\% of their area, the maximum particle
size is of order 2 cm.
As the fine dust and ice are removed, a `mantle' or `crust' of larger
particles dominates the outer layers, and the fraction of the
surface with exposed ice decreases. 
One then expects that short-period comets, which have been resident 
the inner Solar System for multiple orbits, will have 
surface physical properties significantly different
from Trans-Neptunian Objects (TNOs)
or from dynamically new comets from the Oort cloud.
The albedos of small TNOs (including Centaurs) are much higher than those
of short-period comets \citep{kboalbedo}, which supports the model of rapid
surface evolution.

New insight into the nature of comets was provided by the discovery of cometary debris
trails \citep{sykes86}. Unlike the small-particle tails apparent at visible
wavelengths, debris trails consist of large (mm-cm), dark, low-velocity particles more
readily observable at thermal infrared wavelengths. A survey of trails detected 
serendipitously by the
{\it Infrared Astronomical Satellite} concluded that cometary mass loss was primarily in the form
of these large particles and that comets therefore had a greater refractory or rocky
component than previously thought, with a dust/gas mass ratio of 
$\sim ~3/1$, consistent with a
formation location similar to Pluto and Triton \citep{SykesWalker}. One of the
best-observed comets of all time, C/Hale-Bopp, yielded a wealth of new observational
data, including submillimeter photometry suggesting large particles dominate the mass of
the coma of C/Hale-Bopp, with dust/gas $> 5$ \citep{jewittmatthews}.

It was also
recognized that these trails are the first stage in the evolution of a meteor stream.
While dust trails were immediately recognized as the  `genetic link'  between comets and
meteor streams \citep{sykes86}, many meteor streams have been associated with comets
since the mid-nineteenth century, when their radiants were measured and found to coincide
with known comet orbits (Yeomans 1991). For more historical details, we refer the
reader to an excellent review of this field by \citet{jenniskens}.
Kresak (1993) showed that
debris trails would actually produce meteor {\it storms} 
(analogous to the well-known Leonids) if they intersected
the Earth's orbit: the width of debris trails matches the duration
of meteor storms (less than an hour) but is less than that of showers 
(days),
and the particle density within debris trails better matches the extremely
high zenith hourly rate of meteor storms rather than the
order-of-magnitude-lower meteor rate in meteor streams.
Meteor storms also appear to be associated with relatively recent
cometary events, and their orbits match the current orbits of associated
comets. Meteor
streams, on the other hand, significantly deviate in mean orbit from 
that of their
associated comet, as well as having a wide dispersion.
The Taurid meteor complex is widely associated with emissions
from comet 2P/Encke, not in its present orbit, but rather 
that of $10^{4-5}$ yr ago \citep{SteelAsherClube,JonesTaurid}.

Spacecraft with cameras and dust sensors
have now encountered five comets:
{\it Giotto} passed 600 km and {\it Vega} passed 8000 km from 1P/Halley in 1986 \citep{mcdonnellHalley,mazetsHalley};
{\it Giotto} subsequently passed within 100-200 km of Grigg-Skjellerup in 1992 \citep{mcdonnellGrigg};
{\it Deep Space 1} passed 2100 km from 19P/Borelly in 2001 \citep{soderblom};
{\it Stardust} passed 236 km from 81P/Wild 2 in 2004 \citep{brownlee04}; and
then {\it Deep Impact} met 9P/Tempel 1 in 2005 \citep{ahearnDeepImpact}.
The size distribution of detected particles indicates most of
the mass of cometary dust is in the largest particles. Thus, nearly
all estimates of cometary dust mass loss 
based on ground observations at visible wavelengths
are lower limits to the 
actual dust mass loss.
Particularly important, then, is the size of the largest particles.
A particle of order 5 mg struck the front bumper of {\it Giotto}
as it approached comet Halley; the particle perturbed the spacecraft
attitude, generated an ion cloud detected by the spacecraft,
and damaged the spacecraft's star tracker \citep{goldstein}.
During the flyby of Grigg-Skjellerup, the Radio Science Experiment
detected a particle of at least 30 mg \citep{patzold}.
During the recent {\it Stardust} encounter with
81P/Wild 2, there is evidence of a 14 mg particle on the dust
shields \citep{wild2bigparticle}. 
The dust flux monitor yielded a large-particle-dominated
size distribution with cumulative mass index 0.75; these
results also led to the prediction that the dust collector
may contain a mm-size particle, which will dominate the total
collected mass \citep{greenWild2}. 
As the Deep Impact impactor spacecraft approached 9P/Tempel 1,
large particles caused dramatic attitude jogs \citep{ahearnDeepImpact}. 
Such particles would all be bright
meteors if they were to enter the Earth's atmosphere.
Large enough to be only slightly affected
by radiation pressure, they follow orbits similar to their
parent nucleus and only gradually drift away. The estimated dust/gas
mass loss ratio is $>3$ from these impact observations, 
consistent with the value inferred from remote observations of debris
trails by {\it IRAS}, suggesting
that comets are mostly composed of rocky material. 
Furthermore, the dust-to-gas mass ratio of ejected 
material underestimates that of the nucleus, since it does not include
objects too large to be lifted off the nucleus
(unless there are simply no particles larger than ~1 g within the nucleus).

Debris trails provide a unique record of comet emission history in addition to providing
insights into their composition. Over the eight trails detected by IRAS associated with
known short-period comets, \citet{SykesWalker} 
determined the ages of the oldest detected particles to be from years to 
centuries. They inferred that all short-period comets should have trails, and while the
median dust to gas mass ratio for the comets studied was 3, individual values ranged from
1.2 to 4.6 - indicating perhaps a significant variation in composition from comet to comet.
In this paper we report on the first survey of short-period comets since {\it IRAS} made
its observations more than 20 years ago, focusing on the large particle emissions of these objects. Taking
advantage of the greater sensitivity, spatial resolution and pointing capability of the new
{\it Spitzer Space Telescope}, we have quadrupled the detections of short-period comets and
have been able to probe the similarity of the large particle emissions of short-period
comets in more detail than has been possible previously. 


\section{Observations}

\subsection{Methods of observation and data reduction}

All observations presented here were performed with the 
Multiband Imaging Photometer for Spitzer \citep[MIPS;][]{rieke} 
at 24 $\mu$m. The dates and viewing geometries
are summarized in Table~\ref{trailtab}. The observed comets were selected
to be all those short-period (Jupiter-Family) comets
in the inner Solar System ($r<3.5$ AU) visible to {\it Spitzer}
during the time period from 2004 Jan to 2005 May, corresponding to
the initial guaranteed-time and first general-observation cycles
on the observatory; \ncomobs\ comets met these criteria.

All comets were observed using the small-field photometry mode of MIPS,
wherein each $6'\times 6'$ field is observed using a set of 2 telescope
positions with 7 scan-mirror dithers at each. The first set
of observations (PID 210; up to 9/21/2004, excluding 141P), executed during
guaranteed time, covered only 3 fields per comet, with offsets
in telescope coordinates (roughly, ecliptic longitude) centered on
the nucleus, leading and trailing the nucleus by one field of view. 
For the remaining observations executed during general
observing time (PID 20039), maps were built using $6'$ tiles designed
to follow each comet's orbit to at least $10^6$ km following the nucleus
and $10'$ leading the nucleus. Thus a wider field on the sky, following
the nucleus, was observed for the nearby comets.

The basic calibrated data (individual $128\times 128$ pixel calibrated
array images) were improved as follows. We noted a significant
gradient across the individual images; the gradient is strongest
at the beginning of a scan-mirror-driven dither pattern and gradually
decreases. To reduce the effect, we median-combined all images from our 
project at each of the 7 scan-mirror positions (and for each exposure time), 
and subtracted these bias corrections from each original image. This
correction is imperfect due to the prevalence of bright cometary emission
(even after excluding the brightest comets), but improved the image
quality significantly. To ensure that the bias subtraction did not
remove some of the trail brightness, we inspected each bias image to
determine the amplitude of structure, on $<1'$ scales, is
$<0.05$ MJy~sr$^{-1}$. (A large-scale gradient is present, but it
has no bearing on our results as we will subtract a large-scale
gradient from the final mosaics.) The mosaics were composed of 
14 independent combinations of bias-subtracted images, so the
amplitude of bias-image structure that survives into the mosaics
$<0.01$ MJy~sr$^{-1}$. Further, there was no evidence of `trail'
like structure (by which we mean the structures presented
in the images in this paper) in the bias images.

The basic calibrated data were then shifted to each comet's rest frame
by shifting each image to counter the motion of the
comet between the time an image was taken and the first image
of the sequence. The images were mosaiced using
the SSC tool {\it mopex}, which matches the backgrounds in overlapping
portions of images and combines the images into a single celestial grid
with outlier rejection to remove cosmic rays and galactic
protons \citep{mopexref}.

\def\colhead{}

\begin{table*}[h]
\notsotiny
  \caption{Spitzer Comet Trail Survey Summary\label{trailtab}.}
  \begin{tabular*}{\tblwidth}{@{} lrrrccccccccccl@{} }
   \toprule
\colhead{Comet}      & \colhead{Date}    & \colhead{UT}    & \colhead{$T-T_p$} & \colhead{$R$}  & \colhead{$\Delta$} & 
	\colhead{q} & \colhead{e} & \colhead{$T_J$} & \colhead{$T_{pert}$$^a$} &
	\colhead{PA$_{Sun}$} & \colhead{PA$_{trail}$} &
      \colhead{$F_\nu$$^v$}  & 
	\colhead{$dM_3/dt$$^c$} & \colhead{Trail type$^d$} \\
    &         & (hh:mm) & (days) & (AU) & (AU)     & (AU) & & & & ($^\circ$) & ($^\circ$) & (mJy) & (kg~s$^{-1}$) & 
\\
\midrule
2P/Encke              & 06/20/04 & 18:35 & 173 & 2.53 & 2.02 & 0.339 & 0.847& 3.02& $<$1769 &  70.4 & 211.1 & 69   & 26  & long L F \\
4P/Faye                & 10/02/2006 &  2:29  & -44 & 1.73 & 0.98 & 1.67 & 0.567 & 2.75 & 1816 &  75.9 & 264.2  &   7400 & 1       & narrow L F\\
9P/Tempel 1           & 03/14/04 & 15:02 &-478 & 3.75 & 3.40 & 1.502 & 0.633& 2.97& $<$1839 & 272.6 & 259.3 & 0.77 & 14 & common F \\
10P/Tempel 2          & 07/08/04 & 10:12 &-222 & 2.49 & 2.34 & 1.468 & 0.526& 2.96& $<$1846 & 292.4 & 287.8 & 0.60 & 26 & long L F \\
32P/Comas Sola        & 01/26/05 & 14:28 & -66 & 1.93 & 1.36 & 1.834 & 0.569& 2.67& 1912 & 253.1 & 233.6 & 1811 & 0.8 & narrow F \\
36P/Whipple           & 01/22/04 & 10:43 & 199 & 3.24 & 2.79 & 3.088 & 0.259& 2.95& 1922 & 251.4 & 261.6 & 30.2 & 8.1 & intermed F \\

42P/Neujmin 3         & 10/13/04 & 17:25 &  89 & 2.17 & 1.54 & 2.018 & 0.584& 2.63& 1850 & 254.7 & 258.9 & 120  &... & none \\
48P/Johnson           & 10/13/04 & 21:12 &   1 & 2.31 & 1.94 & 2.310 & 0.365& 2.94& $<$1903 & 256.0 & 274.1 & 533  & 2.6 & narrow L F\\
49P/Arend-Rigaux      & 12/05/04 & 13:40 & -82 & 1.64 & 0.99 & 1.367 & 0.612& 2.71& $<$1801 & 236.4 & 270.0 & 1497 &...  & none\\
53P/van Biesbroeck    & 08/06/04 & 12:26 & 302 & 3.94 & 3.30 & 2.415 & 0.551& 2.65& 1850 &  69.8 & 252.5 & 154  &...  & (tail) \\

56P/Slaughter-Burnham & 12/02/04 & 03:47 & -44 & 2.56 & 1.98 & 2.530 & 0.504& 2.71& $<$1800 & 248.0 & 236.7 & 217  & 1.8 & narrow L F \\
62P/Tsuchinshan 1     & 02/03/05 & 21:50 &  57 & 1.61 & 0.96 & 1.491 & 0.577& 2.79& 1960 &  99.2 & 300.8 & 929  & 1.0 & narrow F\\
65P/Gunn              & 08/21/04 & 22:58 & 467 & 3.51 & 2.97 & 2.447 & 0.318& 2.99& 1965 &  76.4 & 239.0 & 188  & 11 & long L \\
67P/Chury-Ger         & 02/23/04 & 03:42 & 554 & 4.47 & 4.05 & 1.292 & 0.631& 2.74& 1959 & 106.6 & 296.8 & 8.6  & 31 & narrow L F \\

69P/Taylor	          & 12/03/04 & 15:20 &   2 & 1.94 & 1.38 & 1.943 & 0.467& 2.81& 1925 & 107.0 & 262.5 & 200  & 0.3 & narrow L \\
71P/Clark             & 04/11/05 & 10:54 &-422 & 3.48 & 3.00 & 1.559 & 0.500& 2.99& $<$1945 & 285.2 & 278.2 & 1.8  & 4.9 & debris L F\\
71P/Clark(epoch2)  & 06/09/2006 & 19:18 & 3 & 1.56 & 0.84 & 1.56 & 0.50 & 2.99 & $<$1945 & 77.2 & 269.9 & 5100 & 2  & narrow L\\
78P/Gehrels 2         & 09/21/04 & 15:05 & -36 & 2.03 & 1.32 & 2.004 & 0.463& 2.88& 1911 &  74.8 & 261.8 & 2400 & 1.3 & narrow L \\
88P/Howell            & 08/27/04 & 05:37 & 136 & 1.96 & 1.47 & 1.388 & 0.556& 2.94& 1907 &  74.3 & 243.5 & 704  & 2.3 & narrow L\\
94P/Russell 4         & 03/14/04 & 16:26 & 198 & 2.60 & 2.18 & 2.229 & 0.364& 3.00& $<$1895 &  98.4 & 287.6 & 45.1 & 9.9 & intermed F \\

103P/Hartley 2        & 01/25/05 & 08:25 & 252 & 2.95 & 2.46 & 1.033 & 0.700& 2.64& 1982 & 122.8 & 273.7 & 67.9 &...  & (tail)\\
104P/Kowal 2          & 03/05/05 & 15:29 & 300 & 3.06 & 2.45 & 1.395 & 0.585& 2.80& $<$1894 & 126.9 & 272.4 & 3.4  & 2.1 & intermed F\\
107P/Wilson-Harringto &	07/10/04 & 09:51& 370& 3.3 & 3.10 & 1.001& 0.621& 3.08& $<$1915 &   ...      &  ...           & 2.4  & ... & none\\
108P/Ciffreo         & 10/05/2006 & 14:30 & -284 & 2.97 & 2.35 & 1.71 & 0.542 & 2.77 & $<$1948 & 249.2 & 271.8 &  3.2  & 0.2   & narrow L F\\
111P/Helin-Roman-Croc & 11/09/04 & 22:26 & -45 & 3.48 & 3.01 & 3.477 & 0.140& 3.02& 1976 &  98.0 & 273.9 & 1.4  & 1.8 & narrow F \\
116P/Wild 4           & 06/23/04 & 05:21 & 518 & 3.71 & 3.13 & 2.166 & 0.377& 3.00& 1987 &  71.9 & 246.3 & 28.4 &...  & (tail) \\
120P/Mueller 1        & 09/17/04 & 10:36 & -13 & 2.75 & 2.09 & 2.743 & 0.337& 2.91& 1957 &  69.4 & 246.8 & 14.2 & 2.1 & narrow L F \\
121P/Shoemaker-Holt 2 & 12/31/04 & 05:18 & 120 & 2.75 & 2.17 & 2.651 & 0.338& 2.87& $<$1915 & 102.0 & 282.1 & 398  &  & (tail) \\

123P/West-Hartley     & 01/22/04 & 05:31 &  44 & 2.16 & 1.71 & 2.130 & 0.447& 2.83& $<$1892 & 107.7 & 311.1 & 825  & 0.8 & narrow L \\
127P/Holt-Olmstead    & 02/20/04 & 21:13 & 253 & 2.74 & 2.24 & 2.158 & 0.368& 2.98& $<$1805 & 267.8 & 256.6 & 7.4  & 4.2 & intermed F\\
129P/Shoemaker-Levy 3 & 03/03/05 & 05:37 & -94 & 2.85 & 2.30 & 2.811 & 0.249& 3.02& 1914 & 267.6 & 271.6 & 74   & 4.9 & long L F \\
131P/Mueller 2        & 12/22/04 & 05:35 &   4 & 2.42 & 1.87 & 2.419 & 0.343& 2.98& 1785 & 245.8 & 255.3 & 36   & 0.3 & narrow F\\
133P/Elst-Pizarro    &   04/11/05  & 08:11&  -811& 3.59 & 3.04 & 2.63 & 0.165 &3.18 & $<$1945 & 115.5 & 292.1 & 4.7  & 0.1: & narrow F faint\\
141P/Machholz 2       & 08/22/04 & 09:54 & 189 & 2.53 & 2.21 & 0.749 & 0.751& 2.71& $<$1968 & 281.5 & 252.9 & 2.5  & ...    & debris L F \\
P/2003 S2  	          & 12/03/04 & 16:00 & 453 & 3.59 & 3.03 & 2.457 & 0.358& 2.94& 1850 & 101.7 & 278.2   & 4.8  & 2.6 & narrow F\\
   \bottomrule
  \end{tabular*}
$^a${Year of last significant orbital perturbation}
$^b${Flux within 12.5$"$ aperture centered on nucleus}
$^c${Mass production rate of trail particles assumed to have $\beta=10^{-3}$, $\rho=1$ g~cm$^{-3}$}
$^d${L=leading, F=following nucleus; `long' means the trail extends to the edge of the image; `(tail)' means 
the small-particle dust tail dominates the image preventing debris trail detection; `intermed' means the trail follows the $\beta=10^{-3}$
syndyne more closely than it follows the projected orbit of the nucleus.}

\end{table*}
\clearpage

\section{Dynamics of large particles}

Particles of different size can be distinguished in the images
due to the 
size-sensitive effects of radiation pressure. As a first-order
approximation---which is relatively accurate for large particles
far from the comet but becomes inaccurate for particles that
are small, recently ejected, or close to the comet---we 
neglect the ejection velocity and assume the particles come
into existence with the same orbit as the nucleus. They
immediately experience radiation pressure from sunlight, 
with a $r^{-2}$ dependence identical to solar gravity and
force ratio
\begin{equation}
\beta \equiv \frac{F_{rad}}{F_{grav}} = \kappa \frac{Q_{pr}}{\rho s},
\end{equation}
where $Q_{pr}$ is the effective efficiency of sunlight absorption and
scattering for radial-directed momentum transfer
(assumed to be unity for large particles), $\rho$ is the
particle density in g~cm$^{-3}$, 
and $s$ is the particle radius (in $\mu$m), and
$\kappa = 0.57$ \citep{burnslamysoter}.
Very large particles, with $\beta\rightarrow 0$, will follow 
nearly the same orbit as the nucleus, feeling negligible radiation pressure.
(In detail, they will still deviate, because of the ejection
velocity of the particle and non-gravitational
forces due to asymmetric outgassing acting on the nucleus.)
As discussed above, the largest particles that can be lifted from
the nucleus due to sublimation are of order cm size, 
so $\beta > 5\times 10^{-5}$ for $\rho\leq 1$ g~cm$^{-3}$. The orbits of particles
with such small $\beta$ cannot
be easily distinguished from the nucleus orbit without observations
over long periods.
As the particles separate from the nucleus, the effect of planetary 
perturbations increases and can ultimately result
in dispersal of the trail.
To first order, the largest particles in the trail lie along the
orbit of the nucleus, and we consider the particles that have been
dispersed due to perturbations as part of the more diffuse zodiacal cloud.
Equivalently, in the terminology of meteor science, the largest
particles lie within the same storm or shower, and the particles
that have been dispersed due to perturbations are sporadic meteors.
Nuclear fragments will follow approximately the same orbit, 
with modification due to the initial velocity of splitting, 
non-gravitational forces due to asymmetric outgassing, and planetary
perturbations over time. In general the large debris from recent orbits
will be splayed along the comet's orbit. 

\begin{figure*}[tbh]
    \includegraphics[width=0.99\textwidth]{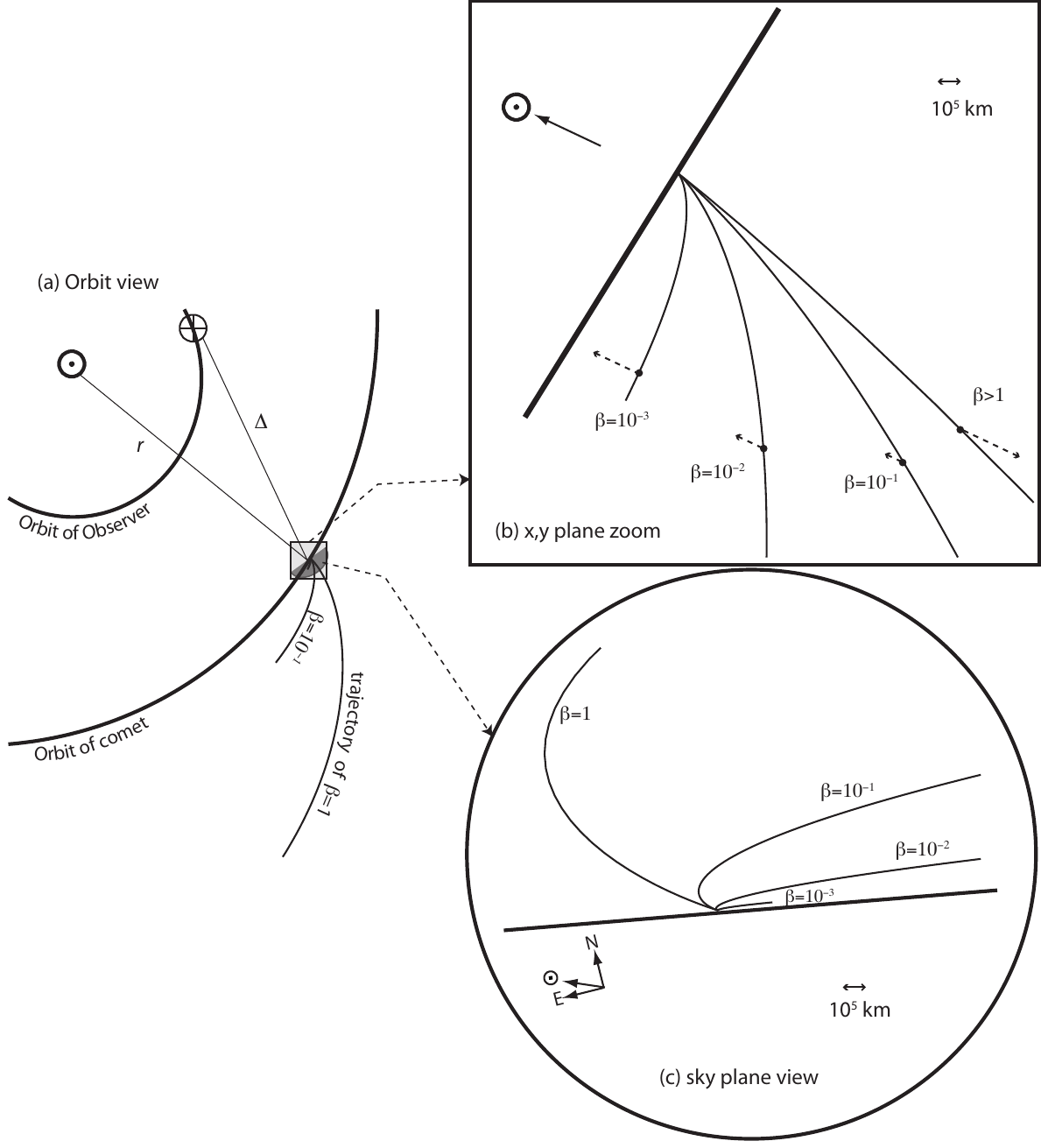}
    \vspace{-1.3in}
\caption{Illustration of cometary debris dynamics and viewing
geometry, drawn specifically for the conditions of the 
observation of 48P/Johnson. 
{\it (a)} In a view looking down onto the ecliptic plane,
the orbits of the observatory and comet are shown together with the 
vectors from comet-to-Sun and comet-to-observatory. The trajectory
of particles with $\beta=1$ (radiation pressure equal to
gravitational force) and $\beta=0.1$ are shown as thin arcs
stretching away from the comet. 
{\it (b)} This is a blow-up of the region in panel {\it (a)} shown
as a light-grey square. The 
trajectories of particles with a range of $\beta$ (ratio of radiation pressure 
to gravity) are shown as thin arcs, and the orbit of the comet as a 
thick line. Dashed vectors on each syndyne show the net direction of
solar forces, which is repulsive for $\beta>1$ and progessively more
attractive as $\beta\rightarrow 0$. 
{\it (c)} This panel shows a view projected onto the plane of the sky,
perpendicular to the observer-comet vector. 
The projected trajectories from panel {\it (b)} are shown; these
are the {\it zero-velocity syndynes}, which serve as a guide to
cometary dust and debris. The smallest particles (largest $\beta$)
feel the most radiation pressure, and they lie closer to the
antisolar direction or $\beta=1$ syndyne. Progressively larger
particles lie closer to the comet's projected orbit, and particles
with $\beta<10^{-3}$ are effectively along the orbit and form
the {\it debris trails} which are the focus of the present survey.
\label{dyntutor}}
\end{figure*}

To track the locations of particles with finite $\beta$, we 
predict their trajectories as follows. Over the two year period
preceding observation, we calculate the detailed motion of
$10^4$ particles produced at even intervals (one 
particle every 2 hr), for each of $\beta=10^{0}$, $10^{-1}$, 
$10^{-2}$, $10^{-3}$, and $10^{-4}$.
We assume each test particle's 
initial velocity and position is the same as the nucleus, including 
non-gravitational effects, if known.
The $x,y,z$ locations of all particles
are then projected onto the sky for the geometry of the {\it Spitzer}
observation. Particles of a given size generally lie along a simple arc
on the sky. This arc of particles with common $\beta$ is 
the {\it zero-velocity syndyne}. The syndynes were generated and
overlaid on each comet image. They trace the
locations and shapes of both the comet tails and trails. 
They do not match the inner cometary comae, which are due to 
recently-ejected, small particles with non-negligible ejection velocity
and some memory of their origin on the rotating nucleus in asymmetrically
located jets.
Figure~\ref{dyntutor} illustrates the zero-velocity syndynes
and their projection on the sky
for one of our survey comets.

To supplement the zero-velocity syndynes, we also generated Monte-Carlo
simulations of particles produced over the 1-year period prior to
observation. The computational techniques are described by
\citet{kelleychury}.
For the Monte-Carlo simulations, $10^6$ particles were
emitted over the 1-year period, with a power-law size distribution 
from $10^{-4}<\beta<10^{-1}$ with mass index $\alpha=-1$, which
distributes mass evenly among particle sizes but favors smaller
particles for surface area.
The velocity distribution is directed toward the Sun, falling of as
the cosine of the zenith angle and with peak velocity 
$v_{ej}=1.0 \sqrt{\beta/r}$ km~s$^{-1}$,
where $r$ is the distance from the Sun at the time of particle ejection.
For typical trail particles produced at $r=1$ AU and with $\beta=10^{-4}$,
the ejection velocity is 10 m~s$^{-1}$. These Monte Carlo simulations are
not adeqaute to trace the debris trails over their entire ages, but rather
they provide a somewhat improved guide to the beginning of the trail,
near the nucleus. Longer-term simulations can follow the evolution
of the debris but are computationally expensive \citep[e.g.][]{vaubaill}.

\begin{figure*}
    \includegraphics[width=0.93\textwidth]{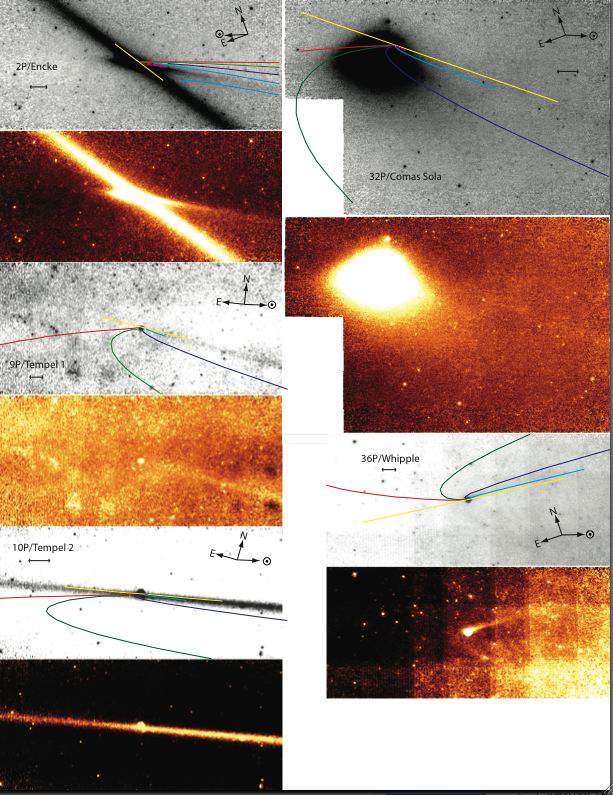}
    \caption{Mid-infrared images and zero-velocity syndynes for comets
2P/Encke, 9P/Tempel 1, 10P/Tempel 2, 32P/Comas Sola, and 36P/Whipple.
For each comet there are two panels. The upper, labeled panel shows the
24 $\mu$m image in greyscale together with the image orientation (celestial
N and E), a scale bar showing $10^5$ km perpendicular to the line of sight,
the projected orbit (yellow line),
and color-coded syndynes that show the location of particles of different
size emitted over the 1 yr period before observation 
(red, green,    blue,    cyan, and magenta correspond to
$\beta=$              1, $10^{-1}$, $10^{-2}$ $10^{-3}$ $10^{-4}$,
where $\beta$ is the ratio of radiation to gravitational force and
is approximately the inverse of the particle size in $\mu$m). For each comet,
the lower panel shows the mid-infrared image alone, in a color table that ranges
from black (faintest) through shades of orange to white (brightest).
\label{comdust1}}
\end{figure*}

\begin{figure*}
    \includegraphics[width=0.999\textwidth]{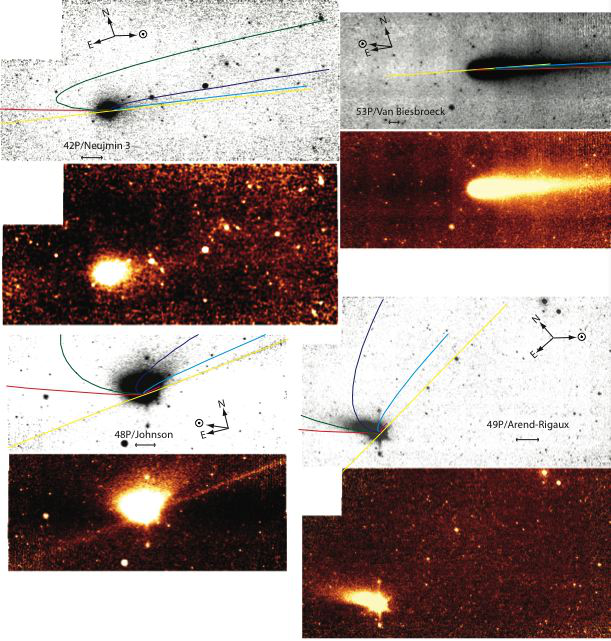}
    \caption{Mid-infrared images and zero-velocity syndynes for comets
42P/Neujmin 3, 48P/Johnson, 53P/van Biesbroeck, and 49P/Arend-Rigaux.
Labels and overlays are the same as in Figure~\ref{comdust1}. The
color images of 42P and 49P have been convolved with a 3-pixel (2.5$''$/pixel)
gaussian.
\label{comdust2}}
\end{figure*}

\begin{figure*}
    \includegraphics[width=0.95\textwidth]{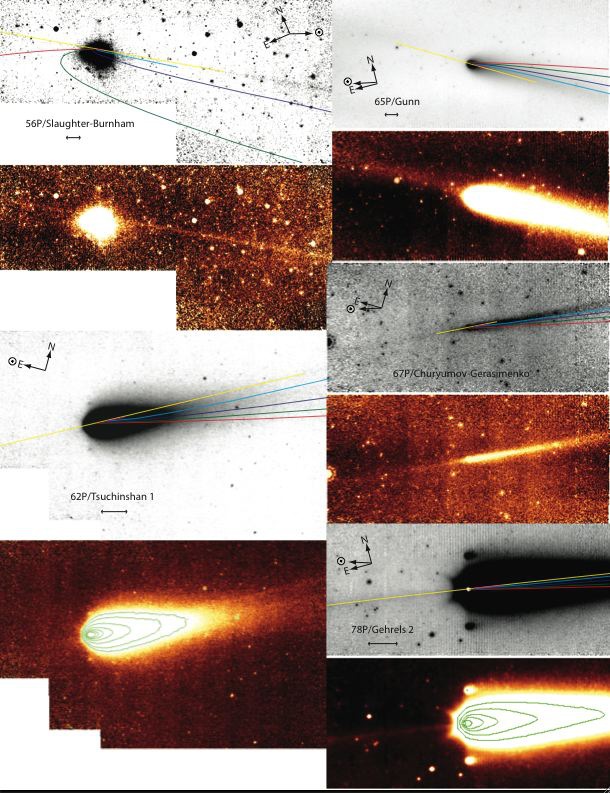}
    \caption{Mid-infrared images and zero-velocity syndynes for comets
56P/Slaughter-Burnham, 62P/Tsuchinshan 1, 65P/Gunn, 
67P/Churyumov-Gerasimenko, and 78P/Gehrels 2.
Labels and overlays are the same as in Figure~\ref{comdust1}. 
For 62P, the color table runs from 37.7--41 MJy~sr$^{-1}$ (linear) and
the contours from 39.5--70 MJy~sr$^{-1}$ (square-root spaced). The
Sun was almost directly E of 62P at the time of observation.
For 78P, the color table runs from 41.1--43.6 MJy~sr$^{-1}$ (linear) and
the contours from 45--100 MJy~sr$^{-1}$ (square-root spaced). 
The two bright splotches 1/3 of the image above and below 78P are latent
images of the nucleus (which saturated the detector) and inner coma.
\label{comdust3}}
\end{figure*}

\begin{figure*}
    \includegraphics[width=0.95\textwidth]{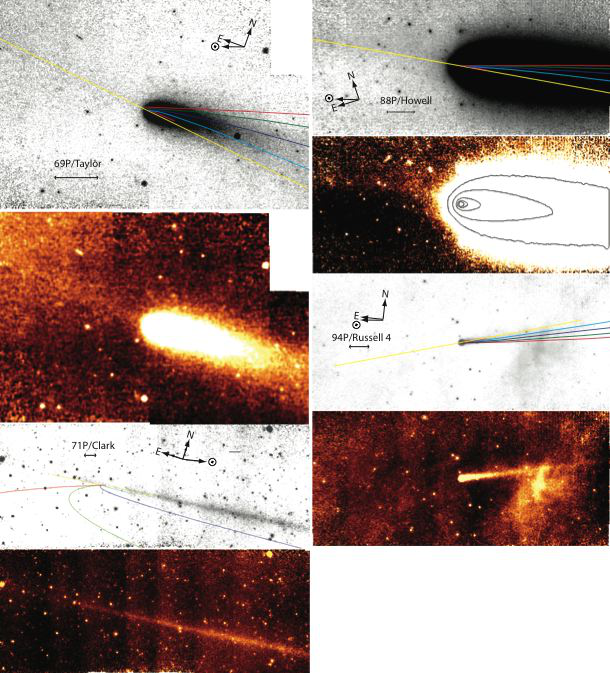}
    \caption{Mid-infrared images and zero-velocity syndynes for comets
69P/Taylor, 71P/Clark, 88P/Howell, and 94P/Russell 4.
Labels and overlays are the same as in Figure~\ref{comdust1}. 
The color image of 69P was convolved with a 2-pixel radius tophat filter.
\label{comdust4}}
\end{figure*}

\begin{figure*}
    \includegraphics[width=0.95\textwidth]{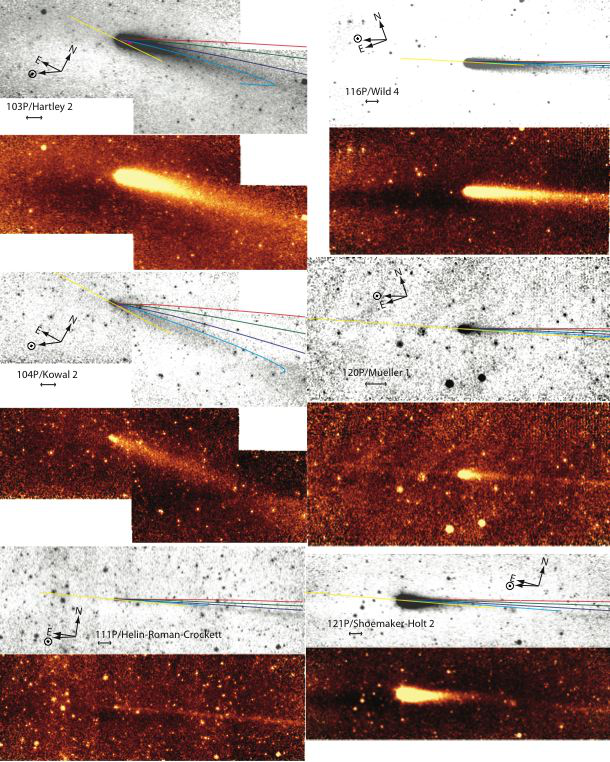}
    \caption{Mid-infrared images and zero-velocity syndynes for comets
103P/Hartley 2, 104P/Kowal 2, 111P/Helin-Roman-Crockett,
116P/Wild 4, 120P/Mueller 1, and 121P/Shoemaker-Holt 2.
Labels and overlays are the same as in Figure~\ref{comdust1}.
\label{comdust5}}
\end{figure*}

\begin{figure*}
    \includegraphics[width=0.95\textwidth]{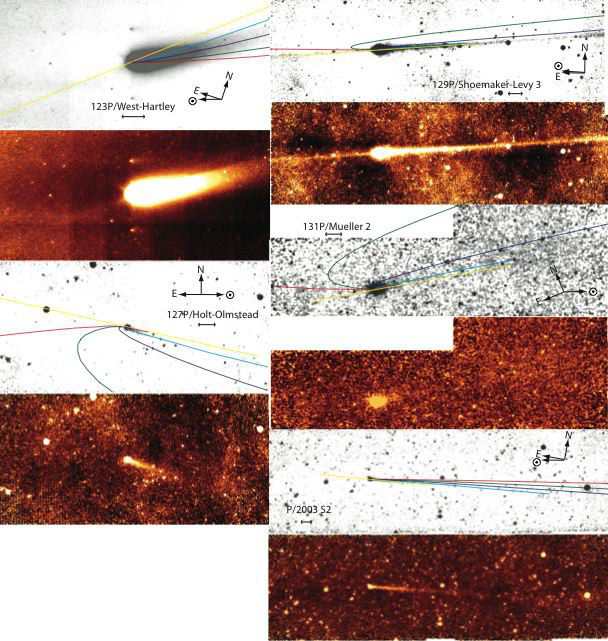}
    \caption{Mid-infrared images and zero-velocity syndynes for comets
123P/West-Hartley, 127P-Holt-Olmstead, 129P/Shoemaker-Levy 3,
131P/Mueller 2, and P/2003 S2.
Labels and overlays are the same as in Figure~\ref{comdust1}. 
For 131P, the images were smoothed with a 2 pixel radius tophat.
\label{comdust6}}
\end{figure*}

\begin{figure*}
    \includegraphics[width=0.8\textwidth]{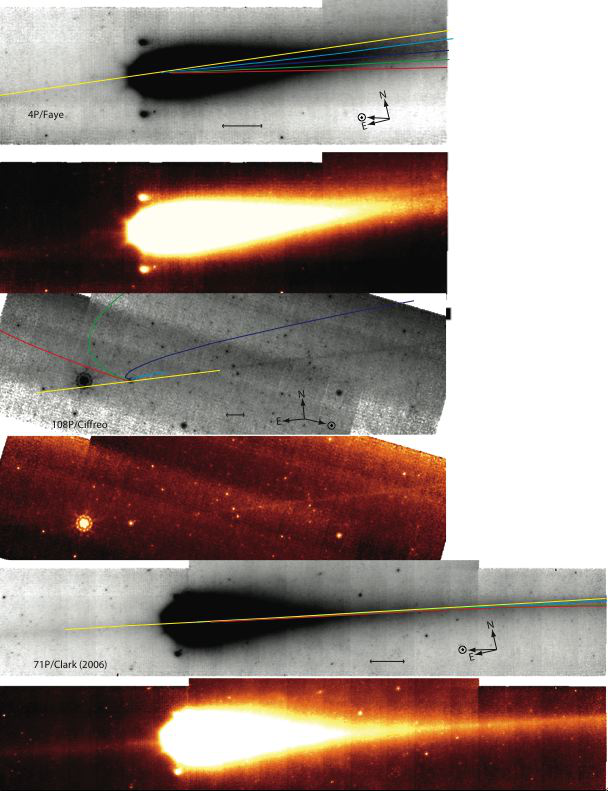}
    \caption{Mid-infrared images and zero-velocity syndynes for comets
4P/Faye. 108P/Ciffreo, and 71P/Clark from 2006 observations.
Labels and overlays are the same as in Figure~\ref{comdust1}. 
\label{comdust7}}
\end{figure*}

\clearpage

\def\cutinhead{}
\def\colhead{}

\begin{table}[h]
  \caption{Debris trail brightness profile fits$^a$\label{proftab}.}
\tiny
  \begin{tabular}{lrcccccc}
   \toprule
\colhead{$\phi$}    & \colhead{$\phi\Delta$}    & \colhead{$I_\nu$}  & \colhead{$\tau$} & 
      \colhead{$W$}  & \colhead{$W\Delta$} &
	\colhead{$F_{1D}$} & \colhead{$dM_3/d\phi$} \\
($''$)    & ($10^4$ km) & (MJy/sr)&($10^{-9}$)& ($''$) &($10^4$ km)& (mJy/$'$) & ($10^{10}$ g/deg) 
\\

\midrule
\multicolumn{5}{l}{2P/Encke}\\
  323    &   47.1      &   0.71  &   5.4 &  54.3 &    7.9    &  54.12     &  135.0 \\
  248    &   36.1      &   0.81  &   6.2 &  54.5 &    8.0    &  62.44     &  155.6 \\
  173    &   25.2      &   0.93  &   7.1 &  45.1 &    6.6    &  59.15     &  147.3 \\
 -128    &  -18.6      &   0.78  &   5.9 &  29.4 &    4.3    &  32.35     &   80.6 \\
 -203    &  -29.6      &   0.87  &   6.6 &  43.0 &    6.3    &  52.69     &  131.4 \\
 -278    &  -40.5      &   0.75  &   5.7 &  42.4 &    6.2    &  44.94     &  111.9 \\
 -353    &  -51.5      &   0.65  &   5.0 &  46.1 &    6.7    &  42.46     &  105.7 \\

\multicolumn{5}{l}{4P/Faye}\\
  446    &   31.8      &   0.27  &   1.2    &  51.4 &    3.7    &  13.9    &  6.5 \\
  295    &   21.1      &   0.24  &   1.1    &  66.1 &    4.7    &  16.0    &  7.7 \\
 
\multicolumn{5}{l}{9P/Tempel 1}\\
 -312    &  -76.7      &   0.14  &   2.1 &  30.1 &    7.4    &   5.79     &   81.8 \\
 -463    & -113.5      &   0.15  &   2.3 &  50.5 &   12.4    &  10.67     &  150.8 \\

\multicolumn{5}{l}{10P/Tempel 2}\\
  475    &   82.4      &   0.35  &   2.6 &  43.2 &    7.5    &  21.58     &   69.9 \\
  400    &   69.4      &   0.32  &   2.4 &  34.7 &    6.0    &  15.88     &   51.4 \\
  325    &   56.4      &   0.34  &   2.5 &  32.8 &    5.7    &  15.60     &   50.5 \\
  250    &   43.4      &   0.39  &   2.8 &  35.4 &    6.1    &  19.21     &   62.3 \\
  175    &   30.4      &   0.39  &   2.9 &  28.3 &    4.9    &  15.41     &   50.0 \\
  100    &   17.4      &   0.47  &   3.5 &  26.6 &    4.6    &  17.58     &   57.0 \\
 -125    &  -21.7      &   0.62  &   4.6 &  32.0 &    5.6    &  27.94     &   90.5 \\
 -200    &  -34.7      &   0.65  &   4.8 &  31.7 &    5.5    &  28.81     &   93.5 \\
 -275    &  -47.7      &   0.64  &   4.8 &  28.5 &    4.9    &  25.86     &   83.9 \\
 -350    &  -60.8      &   0.67  &   5.0 &  30.0 &    5.2    &  28.49     &   92.4 \\
 -425    &  -73.8      &   0.70  &   5.2 &  36.9 &    6.4    &  36.45     &  118.0 \\
 -500    &  -86.8      &   0.73  &   5.4 &  38.6 &    6.7    &  39.83     &  129.0 \\

\multicolumn{5}{l}{32P/Comas Sola}\\
 -543    &  -53.3      &   0.05  &   0.2 &  18.6 &    1.8    &   1.22     &    0.9 \\
 -992    &  -97.5      &   0.06  &   0.3 &  35.4 &    3.5    &   2.96     &    2.2 \\

\multicolumn{5}{l}{36P/Whipple$^b$}\\
 -123    &  -24.7      &   0.12  &   1.4 &  50.8 &   10.3    &   8.59     &   62.4 \\
 -197    &  -39.9      &   0.09  &   1.0 &  34.4 &    6.9    &   4.19     &   30.3 \\
 -273    &  -55.1      &   0.08  &   1.0 &  34.3 &    6.9    &   4.07     &   29.5 \\
 -350    &  -70.7      &   0.08  &   1.0 &  49.7 &   10.0    &   5.87     &   42.5 \\

\multicolumn{5}{l}{48P/Johnson}\\
  400    &  53.4       &   0.06  &   0.4 &  22.2 &    3.0    &   1.86     &    3.7 \\
  325    &  43.4       &   0.07  &   0.4 &  35.5 &    4.7    &   3.39     &    6.7 \\
  250    &  33.3       &   0.08  &   0.5 &  33.0 &    4.4    &   3.73     &    7.4 \\
  175    &   23.3      &   0.09  &   0.6 &  30.3 &    4.0    &   3.86     &    7.7 \\
 -125    & -16.7       &   0.09  &   0.6 &  15.3 &    2.0    &   1.89     &    3.7 \\
 -200    & -26.7       &   0.13  &   0.8 &  24.3 &    3.2    &   4.41     &    8.7 \\
 -275    & -36.7       &   0.15  &   1.0 &  13.6 &    1.8    &   2.79     &    5.5 \\
 -350    & -46.7       &   0.13  &   0.9 &  21.5 &    2.9    &   4.03     &    7.9 \\
 -425    &  -56.7      &   0.19  &   1.2 &  16.2 &    2.2    &   4.27     &    8.4 \\
 -500    &  -66.7      &   0.16  &   1.1 &  15.8 &    2.1    &   3.62     &    7.2 \\

\multicolumn{5}{l}{56P/Slaughter-Burnham}\\
  317    &   45.6      &   0.03  &   0.3 &  27.4 &    3.9    &   1.35     &    3.3 \\
 -282    &  -40.6      &   0.04  &   0.3 &  27.9 &    4.0    &   1.59     &    3.9 \\
 -432    &  -62.2      &   0.04  &   0.3 &  15.1 &    2.2    &   0.91     &    2.2 \\
 -582    &  -83.7      &   0.05  &   0.4 &  21.8 &    3.1    &   1.61     &    3.9 \\
 -732    & -105.3      &   0.07  &   0.5 &  33.4 &    4.8    &   3.07     &    7.4 \\
 -882    & -126.8      &   0.05  &   0.4 &  27.7 &    4.0    &   2.11     &    5.1 \\
-1032    & -148.4      &   0.03  &   0.3 &  15.8 &    2.3    &   0.76     &    1.8 \\

\multicolumn{5}{l}{62P/Tsuchinshan 1}\\
  148    &   10.1      &   0.07  &   0.3 &  78.0 &    5.3    &   7.78     &    2.2 \\

\multicolumn{5}{l}{65P/Gunn}\\
  395    &   82.9      &   0.09  &   1.3 &  35.9 &    7.5    &   4.80     &   45.9 \\
  320    &   67.1      &   0.14  &   1.9 &  51.4 &   10.8    &  10.00     &   95.3 \\
  245    &   51.5      &   0.10  &   1.4 &  39.7 &    8.3    &   5.70     &   54.3 \\

\multicolumn{5}{l}{67P/Churyumov-Gerasimenko}\\
   400    &  117.9	&   0.05  &   1.1 &  25.6 &    7.5	&   1.88	 &   54.1 \\
   325    &   95.8	&   0.05  &   1.2 &  30.3 &    8.9	&   2.28	 &   65.5 \\
   250    &   73.7	&   0.05  &   1.1 &  29.7 &    8.8	&   2.07	 &   59.2 \\
   175    &   51.6	&   0.07  &   1.5 &  27.4 &    8.1	&   2.73	 &   78.6 \\
  -200    &  -59.0	&   0.39  &   8.6 &  25.2 &    7.4	&  13.98	 &  401.1 \\
  -275    &  -81.1	&   0.30  &   6.6 &  27.0 &    8.0	&  11.58	 &  332.4 \\
  -350    & -103.2	&   0.27  &   5.8 &  30.6 &    9.0	&  11.49	 &  329.5 \\
  -425    & -125.3	&   0.22  &   4.8 &  35.7 &   10.5	&  11.03	 &  316.8 \\
  -500    & -147.4	&   0.21  &   4.5 &  51.5 &   15.2	&  15.01	 &  431.1 \\

\multicolumn{5}{l}{69P/Taylor}\\
  583    &   57.9	    &   0.04  &   0.2 &  13.8 &    1.4	&   0.85	 &    0.7 \\
  357    &   35.5	    &   0.04  &   0.2 &  17.0 &    1.7	&   0.97	 &    0.8 \\

\end{tabular}
$^a${
$\phi$ is the distance behind the nucleus at which the trail profile was analyzed.
$F_{1D}$ is the brightness integrated perpendicular to the trail.
$dM_3/d\phi$ is the mass per unit projected angle on the sky.}
$^b${Profiles taken through map rotated at angle of $\beta=10^{-3}$ syndyne rather than 
$\beta=0$, to match observed orientation of trail.}
\end{table}

\begin{table}[h]\ContinuedFloat
  \caption{Debris trail brightness profile fits (continued).}
\tiny
  \begin{tabular}{lrcccccc}
   \toprule
\colhead{$\phi$}    & \colhead{$\phi\Delta$}    & \colhead{$I_\nu$}  & \colhead{$\tau$} & 
      \colhead{$W$}  & \colhead{$W\Delta$} &
	\colhead{$F_{1D}$} & \colhead{$dM_3/d\phi$} \\
($''$)    & ($10^4$ km) & (MJy/sr)&($10^{-9}$)& ($''$) &($10^4$ km)& (mJy/$'$) & ($10^{10}$ g/deg) 
\\

\midrule
\multicolumn{5}{l}{71P/Clark}\\
    90    &   19.6      &   0.07  &   0.9 &  20.9 &    4.5    &   2.10     &   20.1 \\
   -60    &  -13.0      &   0.07  &   1.0 &  22.7 &    4.9    &   2.37     &   22.7 \\
  -135    &  -29.3      &   0.08  &   1.1 &  31.1 &    6.8    &   3.47     &   33.3 \\
  -285    &  -61.9      &   0.12  &   1.5 &  27.3 &    5.9    &   4.44     &   42.5 \\
  -360    &  -78.2      &   0.10  &   1.3 &  28.5 &    6.2    &   4.00     &   38.2 \\
  -435    &  -94.5      &   0.10  &   1.3 &  26.9 &    5.8    &   3.86     &   36.9 \\
  -510    & -110.8      &   0.11  &   1.5 &  29.6 &    6.4    &   4.77     &   45.5 \\
  -585    & -127.1	&   0.12  &   1.5 &  28.5 &    6.2	&   4.64	 &   44.5 \\
  -660    & -143.4	&   0.11  &   1.4 &  32.2 &    7.0	&   4.94	 &   47.3 \\
  -735    & -159.7	&   0.13  &   1.7 &  26.1 &    5.7	&   4.85	 &   46.3 \\
  -810    & -176.0	&   0.10  &   1.4 &  33.3 &    7.2	&   4.91	 &   47.2 \\

\multicolumn{5}{l}{78P/Gehrels 2}\\
  323    &   31.0      &   0.18  &   1.0 &  19.5 &    1.9    &   4.92     &    3.7 \\
  285    &   27.4      &   0.16  &   0.9 &  11.7 &    1.1    &   2.63     &    2.0 \\
  135    &   13.0      &   0.18  &   0.9 &  18.7 &    1.8    &   4.64     &    3.5 \\

\multicolumn{5}{l}{88P/Howell}\\
  437    &   44.8      &   0.12  &   0.6 &  47.2 &    4.8    &   7.91     &    7.0 \\
  288    &   29.5      &   0.11  &   0.6 &  27.1 &    2.8    &   4.25     &    3.8 \\

\multicolumn{5}{l}{94P/Russell 4$^b$}\\
 -125    &  -20.0      &   0.46  &   3.7 &  22.1 &    3.5    &  14.41     &   43.3 \\
 -275    &  -43.9      &   0.21  &   1.7 &  28.5 &    4.5    &   8.60     &   25.9 \\
 -500    &  -79.8      &   0.09  &   0.7 &  18.3 &    2.9    &   2.23     &    6.7 \\

\multicolumn{5}{l}{104P/Kowal 2$^b$}\\
 -223    &  -39.6      &   0.06  &   0.6 &  31.7 &    5.6    &   2.50     &   12.6 \\
 -373    &  -66.3      &   0.03  &   0.3 &  24.0 &    4.3    &   1.07     &    5.4 \\

\multicolumn{5}{l}{108P/Ciffreo}\\
 -403    &  -68.7      &   0.05  &   1.33    &  11.1 &    1.9    &   0.8    & 0.1 \\
 -283    &  -48.2      &   0.05  &   1.20    &  11.1 &    1.9    &   0.7    &  0.1\\
 -204    &  -34.8      &   0.07  &   1.71    &  13.2 &    2.3    &   1.3    & 0.1 \\

\multicolumn{5}{l}{111P/Helin-Roman-Crockett}\\
 -170    &  -36.9      &   0.04  &   0.5 &  22.1 &    4.8    &   1.18     &   11.3 \\
 -320    &  -69.5      &   0.05  &   0.7 &  15.8 &    3.4    &   1.10     &   10.7 \\
 -470    & -102.0      &   0.05  &   0.6 &  21.9 &    4.8    &   1.43     &   13.6 \\


\multicolumn{5}{l}{120P/Mueller 1}\\
   443    &   67.1	&   0.04  &   0.3 &  13.3 &    2.0	&   0.76	 &    2.3 \\
   293    &   44.4	&   0.03  &   0.3 &  12.0 &    1.8	&   0.51	 &    1.6 \\
   143    &   21.6	&   0.02  &   0.2 &  12.3 &    1.9	&   0.37	 &    1.1 \\
  -120    &  -18.2	&   0.11  &   0.9 &  20.6 &    3.1	&   3.12	 &    9.5 \\
  -195    &  -29.6	&   0.06  &   0.6 &  13.9 &    2.1	&   1.25	 &    3.8 \\
  -270    &  -40.9	&   0.04  &   0.3 &  21.2 &    3.2	&   1.20	 &    3.7 \\
  -345    &  -52.3	&   0.08  &   0.7 &  20.2 &    3.1	&   2.39	 &    7.3 \\
  -420    &  -63.7	&   0.07  &   0.7 &  16.5 &    2.5	&   1.74	 &    5.3 \\


\multicolumn{5}{l}{123P/West-Hartley}\\
  425    &   52.8      &   0.04  &   0.2 &  25.4 &    3.2    &   1.46     &    2.0 \\

\multicolumn{5}{l}{127P/Holt-Olmstead$^b$}\\
 -127    &  -20.7      &   0.10  &   0.9 &  40.7 &    6.6    &   5.71     &   19.9 \\
 -203    &  -33.0      &   0.06  &   0.5 &  31.6 &    5.1    &   2.59     &    9.1 \\

\multicolumn{5}{l}{129P/Shoemaker-Levy 3}\\
  430    &   71.8      &   0.14  &   1.3 &  31.2 &    5.2    &   5.96     &   23.4 \\
  355    &   59.3      &   0.12  &   1.1 &  32.3 &    5.4    &   5.42     &   21.2 \\
  280    &   46.7      &   0.18  &   1.6 &  21.5 &    3.6    &   5.33     &   21.0 \\
  205    &   34.2      &   0.16  &   1.5 &  23.4 &    3.9    &   5.39     &   21.1 \\
  130    &   21.7      &   0.16  &   1.5 &  20.0 &    3.3    &   4.64     &   18.2 \\
 -170    &  -28.4      &   0.37  &   5.77    &  21.9 &    3.7    &  11.52    &  25.23  \\
 -245    &  -40.9      &   0.38  &   5.91    &  19.4 &    3.2    &  10.42    &  22.81  \\
 -320    &  -53.4      &   0.36  &   5.57    &  19.4 &    3.2    &   9.85    &  21.58  \\
 -395    &  -65.9      &   0.30  &   4.66    &  20.8 &    3.5    &   8.82    &  19.32  \\
 -470    &  -78.5      &   0.25  &   3.90    &  19.9 &    3.3    &   7.06    &  15.45  \\
 -545    &  -91.0      &   0.24  &   3.66    &  23.8 &    4.0    &   7.93    &  17.36  \\
 -620    & -103.5      &   0.24  &   3.77    &  27.8 &    4.6    &   9.55    &  20.92  \\
 -695    & -116.0      &   0.25  &   3.86    &  27.0 &    4.5    &   9.49    &  20.78  \\
 -770    & -128.5      &   0.23  &   3.63    &  22.9 &    3.8    &   7.59    &  16.63  \\
 -845    & -141.1      &   0.22  &   3.39    &  25.7 &    4.3    &   7.92    &  17.35  \\
 -920    & -153.6      &   0.19  &   3.01    &  27.5 &    4.6    &   7.54    &  16.52  \\
 -995    & -166.1      &   0.21  &   3.30    &  35.7 &    6.0    &  10.73    &  23.51  \\
-1070    & -178.6      &   0.21  &   3.21    &  33.5 &    5.6    &   9.81    &  21.48  \\

\multicolumn{5}{l}{131P/Mueller 2}\\
 -318    &  -43.2      &   0.04  &   0.3 &   6.5 &    0.9    &   0.40     &    0.8 \\

\cutinhead{P/2003 S2}\\
  -183    &  -40.1     &   0.09  &   1.2 &  16.7 &    3.7    &   2.06     &   21.3 \\
  -258    &  -56.6     &   0.04  &   0.5 &  15.5 &    3.4    &   0.82     &    8.4 \\
  -333    &  -73.0     &   0.04  &   0.5 &   9.1 &    2.0    &   0.46     &    4.7 \\
  -408    &  -89.5     &   0.03  &   0.4 &  17.1 &    3.7    &   0.75     &    7.8 \\
\end{tabular}
\end{table}

\section{Measured debris trail properties}

\def\arcsec{$''$}

For each debris trail, we measured the brightness and width 
in one-dimensional slices perpendicular
to the orbit (coadding along the orbit by 30-90 pixels [2.5$''$/pixel] 
depending on signal-to-noise).
Table~\ref{proftab} shows the slice locations
in apparent angular distance from the nucleus, $\phi$, and physical distance 
$\phi\Delta$, along the orbit projected on the sky, where $\Delta$ is 
the distance to the observatory at the time of observation.
For each profile, a Gaussian fit was made, and Table~\ref{proftab} lists the
peak surface brightness $I_\nu$ (median 0.12 MJy~sr$^{-1}$,
range 0.02-0.9 MJy~sr$^{-1}$) and full-width-at-half-maximum in angular units $W$ 
(median 27\arcsec, range 7-78\arcsec) and
physical units $W\delta$ (median $5\times 10^5$ km,
range 0.9-15 $\times 10^5$ km),
in the viewing plane perpendicular to the line of sight.
When fitting the Gaussians to the slices, a polynomial was simultaneously
fitted to portions of the slice on either side of the trail.
The optical depth, $\tau$, was derived by assuming the grains are
blackbody emitters with temperature $T=300 r^{-1/2}$ K,
as was measured for trails with multi-band IRAS detections
\citep{sykescometbook}. 
Note this temperature is significantly warmer than that obtained 
by dust (grey, isothermal particles), for which the temperature at 1 AU
$T_1< 278$ K. As discussed by
\citet{sykescometbook}, trail particles are more consistent with 
rapidly rotating, randomly oriented, zero-albedo 
particles maintaining a latitudinal temperature variation
across their surfaces.
Table~\ref{proftab} lists the resulting $\tau$ (median $2\times 10^{-9}$,
range 0.3-16 $\times 10^{-9}$).
The 1-dimensional flux of the trail, $F_{1D}=I_\nu W$, measures
the flux per unit length along the trail; Table~\ref{proftab}
lists $F_{1D}$ (median 4.6 mJy/arcmin, range 0.4-62 mJy/arcmin).

\subsection{Debris age}

First, we estimate the particle ages analytically, 
to determine how the trajectories of debris evolve
as a function of the ratio of radiation pressure to gravity, $\beta$,
the ejection velocity, $v_{ej}$, and the comet's orbit's perihelion
distance and eccentricity, $q$ and $e$.
For particles ejected at perihelion, in the direction of
the comet's motion, with velocity $v_{ej}$, the rate of
separation in mean anomaly of a particle from the nucleus is
\begin{equation}
\frac{d\theta}{dt} = {\sqrt{\frac{GM}{a^3}}} {\frac{2+e}{1-e}} \beta
 + 3 {\frac{v_{ej}}{a}} { \left( \frac{1+e}{1-3} \right) },
\end{equation}
where the first term is due to radiation pressure and the second one
is due to the ejection velocity \citep[eq. 1 and 2 of ][]{SykesWalker}.
If the ejection velocity scales as
\begin{equation}
v_{ej} = v_1 \beta^{1/2} q(AU)^{-1/2} \label{equation:vej}
\end{equation}
\citep[cf.][]{whippleII}
where for example \citet{reachEncke} used $v_1=1$ km~s$^{-1}$ for
2P/Encke, then 
\begin{equation}
\frac{d\theta}{dt} = \sqrt{\frac{GM}{a^3}} {\frac{2+e}{1-e}} \beta
\left[1 + 3 v_1 \left(\frac{GM}{1\,{\rm AU}}\right)^{-1/2}  \beta^{-1/2} 
\frac{\sqrt{1+e}}{2+e}\right].
\end{equation}
For the comets in our sample,
particles with $\beta=10^{-3} \beta_3$
separate from the nucleus at a rate
\begin{equation}
\frac{d\theta}{dt} = 0.27 \left[1 + 1.6 \frac{v_1}{{\rm km~s}^{-1}} \beta_3^{-1/2}\right] 
\beta_3 \,\,{\rm deg~yr}^{-1},
\end{equation}
where $\beta_3\equiv\beta/10^{-3}$,
indicating that the radiation pressure and ejection velocity terms are comparable.
The median $d\theta/dt=0.7'$/yr. 

We can then approximately convert the change in mean anomaly to the separation, $\phi$,
on the sky using (geometrically)
\begin{equation}
\frac{d\phi}{d\theta}=\frac{r}{\Delta} \frac{df}{d\theta},
\label{dphidtheta}
\end{equation}
and the derivative of true anomaly per unit mean anomaly
\begin{equation}
\frac{df}{d\theta} = \frac{1+e}{1-e} \frac{\tan\frac{E}{2}}{\cos^2\frac{E}{2}}
\frac{\cos^2\frac{f}{2}}{\tan\frac{f}{2}} \frac{r}{a}.
\end{equation}
The median value for our observations is $d\phi/d\theta=1.1$ (with a range 0.66 to 2.06).
Combining these equations, the median $d\phi/dt=0.8'$/yr.

\begin{figure}
    \includegraphics[width=0.49\textwidth]{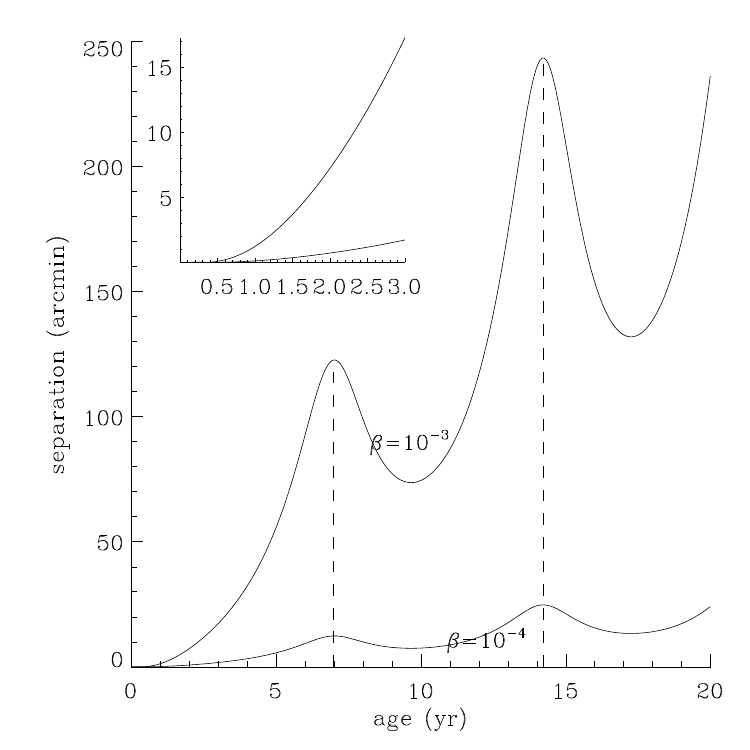}
    \caption{The separation of meteoroids from the nucleus as a function of particle age,
for 129P/Shoemaker-Levy 3 at the epoch and sky-plane of the {\it Spitzer}
observation in 2005. Dashed lines indicate the ages of
particles ejected at each of the previous two perihelion passages.
The inset shows the separation versus time for particles younger than 3 yr, at which time zero-ejection-velocity particles with
$\beta=10^{-3}$ would reach the edge of the {\it Spitzer} image
in Figure~\ref{comdust6}.\label{sepfig}}
\end{figure}

We now estimate the ages of the oldest observed particles
from numerical orbit integration, which should
be more accurate, because fewer approximations are required. The orbits of particles
with $\beta=10^{-3}$ were integrated to the present beginning on an ejection date ranging
from 1 day to 2 years into the past. For these calculations an ejection velocity of
zero and a comet mass of zero are assumed, 
i.e. the particles simply separate from the nucleus according to gravity from the Sun and
planets, and radiation pressure from the Sun.
Their coordinates on the date of the {\it Spitzer}
observation were then compared to those of the nucleus, yielding the separation (on the
projected sky-plane)
from the nucleus as a function of the particle age. This separation of any given
particle increases with time, but particles produced at perihelion separate from
the nucleus more quickly and overtake those produced just earlier. Figure~\ref{sepfig}
shows the separation versus age for 129P/Shoemaker-Levy 3. 
For comets observed pre-perihelion, such as 129P, the separation versus age is
roughly linear for young particles.
The separation is noticeably nonlinear for particles produced near perihelion, 
which makes the trail expansion rate especially nonlinear for comets observed
post-perihelion.
With the notable exception of 2P/Encke, the rate of separation is monotonic over the region
covered by the {\it Spitzer} images, allowing an accurate numerical derivative.
For all comets, we estimate $d\phi/dt$ using the geometric mean of the separation/age
for particles that are $1'$ and $10'$ from the nucleus.
The resulting values have median $d\phi/dt=15'$/yr, with a range 5.7--40$'$/yr.
The numerical results need to be increased to take into account the effect of nonzero
ejection velocity. Based on the equations above, the increase in separation for 
$v_1=1$ km~s$^{-1}$,$\beta=10^{-3}$, and sunward-hemisphere 
ejection, is a factor of 2.6.

The numerical determinations of the rate of separation are consistently higher
than the analytic estimate. Further, the numerical determinations of the separation
rate for $\beta=10^{-3}$ and $\beta=10^{-4}$ do not scale with $\beta$ as the
analytic relations predict (being instead much more shallow).
In order to reach the analytic estimate, some
approximations were made, which apparently led to the difference. We will
use the numerical results from the orbit integrations in this paper.
The trail profiles in Table~\ref{proftab} cover $\phi\sim 10'$ 
from the nuclei.
Thus the observed debris trail particles are 
$\sim 0.6 (10^{-3}/\beta)$ yr old. 
Taking into account the sizes of the {\it Spitzer} images for the various comets,
the range of ages is 0.3--2 yr.
This does not mean that all 
(or even most) of the particles were emitted 0.6 yr ago, but only that
a particle with $\beta\sim 10^{-3}$ emitted 0.6 yr ago would be at the
observed location; particles a factor of 2 smaller or larger in size than 
$\beta=10^{-3}$ (or ejected with velocities smaller or larger) 
would be correspondingly further behind or closer to the nucleus.

In many cases, the debris trails continue to the edges
of the {\it Spitzer} image, so older particles are certainly 
present, as clearly demonstrated by the {\it IRAS} trails 
that extend from more than 1$^\circ$
up to 90$^\circ$ of the comet orbit in mean anomaly, corresponding to 
minimum trail ages of
2.6--140 yr \citep{SykesWalker}. The trail ages exceed the
orbital period for many comets, so a given trail contains debris from
many revolutions. 

\begin{figure}
    \includegraphics[width=0.49\textwidth]{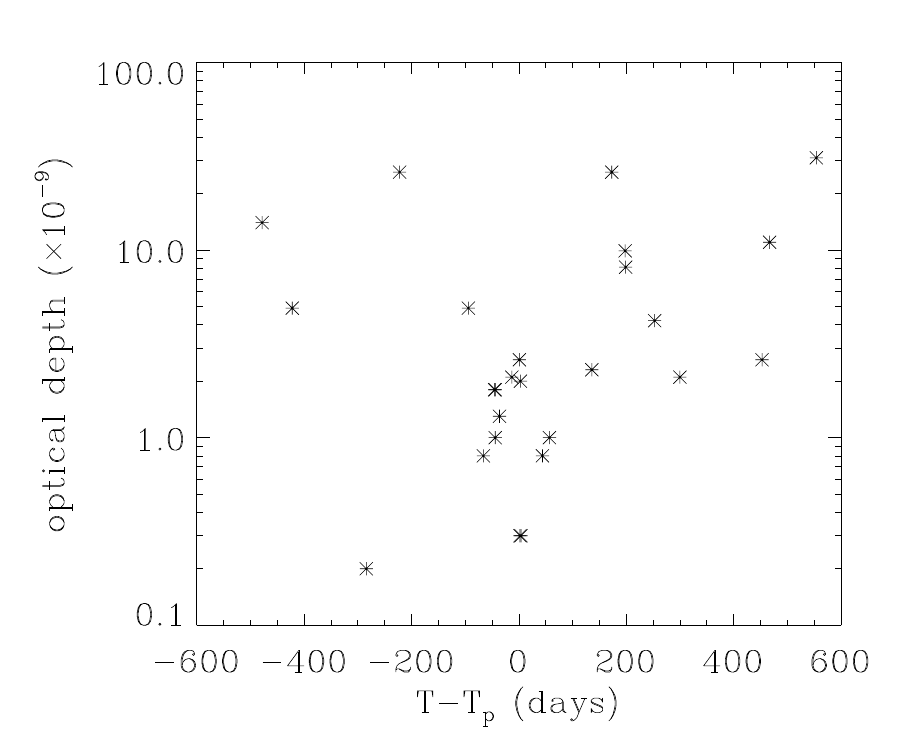}
    \caption{Trail optical depth versus the time of observation with 
respect to the perihelion date, for 30 comets.\label{tmtp}}
\end{figure}

Further evidence of the longevity of debris trail particles is that
trails are present even for comets observed far before perihelion.
Figure~\ref{tmtp} shows that there is no significant
asymmetry in the trail optical depth with respect to perihelion.
While comets show a wide range of behaviors pre- and post-perihelion,
none of them are active {\it only} at aphelion.
Thus, the material observed in a trail on the present revolution 
was largely produced on previous revolutions. This is in contrast to the 
dust tails and comae, which are ephemeral and only due to the present
revolution.
The near-nuclear observations we present here contain
$\beta=10^{-3}$ particles from the present and previous revolution, for comets 
observed post-perihelion, and particles from the previous revolution,
for comets observed pre-perihelion. 
They also contain $\beta\le 10^{-4}$
particles from previous revolutions.

An upper limit on the debris trail age is set by gravitational perturbations.
If debris trails were due only to very large particles, e.g. with $\beta<10^{-5}$,
then trails arise from many revolutions of the comet. 
In this case, the
trail particles would diffuse from the present comets' orbits due to differential
gravitational perturbations. Indeed even minor perturbations and orbital shifts due
to non gravitational and Yarkovsky forces (which will be different for meteoroids
and the nucleus) make it very unlikely to have more than a few revolutions contributing
to the debris trails, which follow the {\it current} orbit of the nucleus very closely.
Instead, the trail would be distorted by minor perturbations and truncated 
by significant perturbations.
Table~\ref{trailtab} shows the date, $T_{pert}$, of the last passage of each comet within 
Jupiter's Laplacian gravitational sphere of influence (0.32 AU)
as calculated by 
K. Kinoshita\footnote{http://www9.ocn.ne.jp/~comet/index.html}.
We see that the date of the last significant perturbation 
by Jupiter was before 1950 for all but 7 comets.
 (62P, 65P, 67P,
103P, 111P, 116P, and 120P), and 
these comets present a range of trail
properties similar to the rest of the sample, including one comet with
a very bright leading+following trail (65P) and one nondetection (103P).
Minor perturbations,
within 1 AU of Jupiter, are much more common; the median time
since such a perturbation for our sample comets is ~\pertdateone.

\citet{SykesWalker} set an independent upper limit on the
trail ages, and a lower limit on $\beta$, by requiring particles ejected 
at perihelion have enough time to reach both the
furthest leading and following extents of the observed trails. 
They found ages of 11--660 yr with a median of 74 yr. These
ages are often greater than the time since the last minor
perturbation (within 1 AU of Jupiter), suggesting such minor perturbations
are not effective at dispersing trails.

\subsection{Debris mass production rate}

The masses of the debris trails are estimated as follows.
A straightforward estimate of the mass per unit projected angle on the
sky, $\phi$, assuming a tapered cylindrical shape for the trail, yields
\begin{equation}
\begin{aligned}
\frac{dM}{d\phi} &=& \frac{\pi}{3} \kappa W \Delta^2 \tau \beta^{-1} \\
 &=& 1.1\times 10^{9} W('') \Delta^2
 \left(\frac{\tau}{10^{-9}}\right) 
 \left(\frac{10^{-3}}{\beta}\right) \\
 && {\rm g/deg.}
\end{aligned}
\end{equation}
Since the brightness profiles are only taken in the debris trail,
only the values of $\beta=10^{-(3-4)}$ that are consistent with the centroid,
width, and length of the trail apply.

We list the mass per unit apparent separation
in Table~\ref{proftab} as $dM_3/d\phi$, for an
assumed particles size $\beta=10^{-3}$.
For many of the images, this is an upper limit to $\beta$,
and therefore $dM_3/d\phi$ is a lower limit to the mass.
The values of $dM_3/d\phi$ are independent of the particle dynamics.
For all the trail profiles in Table~\ref{proftab},
the median value $dM_3/d\phi = 21\times 10^{10}$ g/deg, and
the range is 0.6--430 $\times 10^{10}$ g/deg.

We can now use the particle separation rates (from the
numerical integrations discussed in the previous section),
$d\phi/dt$, together with the observed mass per unit
separation, $dM_3/d\phi$, 
to estimate the mass-loss rates, $dM/dt$,
required to produce the debris trails.
The median $dM/dt=4$ kg~s$^{-1}$, and
the range for all the comet profiles is 0.2--36 kg~s$^{-1}$.
Note that $dM/d\phi \propto\beta^{-1}$ while the analytic estimate 
$d\phi/dt \propto \beta$, so the mass loss rate $dM/dt$ should be
independent of the assumed particle size. However the numerical estimates
of $d\phi/dt$ show much less dependence on $\beta$, with the data in
Figure~\ref{sepfig} yielding $d\phi/dt\propto \beta^{0.3}$ and
suggesting the mass-loss rates scale as $(\beta/10^{-3})^{-0.7}$.

For comparison, using his "inverse tail method" to compare dynamical
models to surface brightness images,
\citet{fulle96} measured as meteoroid mass production rate of 
100 kg~s$^{-1}$
near perihelion for 10P/Tempel 2, and \citet{fulle90} measured meteoroid
mass production rates of 30--50 kg~s$^{-1}$ for 2P/Encke
and 6P/D'Arrest; those results are in general agreement with ours, being
somewhat higher because they refer to dust production closer to perihelion. 
\citet{SykesWalker}, using the {\it IRAS} observations of entire debris trails 
for 8 comets, measured meteoroid mass production rates of 4--250 kg~s$^{-1}$.
For the 5 comets in common between the {\it IRAS} and {\it Spitzer} surveys
(2P, 9P, 10P, 65P, and 67P), the $dM/dt$ inferred in this paper are 
comparable (this paper/SW92 = 0.1, 1.8, 2.3, 0.6, 8, respectively).
The comet for which the mass-loss rates are most discrepant is 2P/Encke.
\citet{reachEncke} used ISOCAM observations and Monte Carlo simulations of 2P/Encke 
 to obtain 200-600 kg~s$^{-1}$ for the 1997 apparition, in
agreement with \citet{SykesWalker}'s value of 260 kg~s$^{-1}$.
The low value for 2P/Encke in the present paper is likely due to observing
only a small portion of the trail, hence sampling a very restricted range of
emission history. The high value for 67P is likely due to a line-of-sight 
enhancement or `neck-line' as discussed below. The overall agreement between
the different observations and calculations in this
paper and \citet{SykesWalker} is encouraging.

Total trail masses are discussed for individual comets in \S\ref{sec:individual}.
Since our {\it Spitzer} images generally cover only limited portions
of the orbits, and many trails clearly extend to the edges of the
images, we can estimate in general only lower limits to the trail
mass. (This effect goes in the same direction [underestimating
the mass] as our conservative assumption of $\beta=10^{-3}$.) 
For the 6 comets with reasonable total mass estimates,
the range is $4\times 10^{10}$ g to $9\times 10^{13}$ g. 
Lower limits for the others range from $>2\times 10^{10}$ g to
$>8\times 10^{11}$ g.
The trail lengths found in the
{\it IRAS} survey were of order degrees (in mean anomaly), 
so the total trail masses from the present survey are of order 
$dM/d\theta$ (with $\theta$ in degrees), 
i.e. typically $10^{11-12}$ g. 
These masses extend to about 1 order
of magnitude lower than those of the {\it IRAS}-detected 
trails \citep{SykesWalker},
due to the greater sensitivity of the present survey.
The masses remain lower limits, because even larger (cm-sized)
particles may be present, and there could be a fainter component of
the debris trails extending over a large fraction of the orbit,
which our observations do not cover.
The existence of meteor showers over large fractions of comets'
orbits testifies to the existence of such an extended population of debris.

\subsection{Comparing meteoroids to ice}

The mass-loss rate in large particles can be compared to that due to
ice sublimation as follows. 
If the H$_2$O production rate is $Q_{{\rm H}_2{O}}$, then
the mass loss rate of sublimating ice is
\begin{equation}
\frac{dM_{ice}}{dt} = 2.9\times 10^2 {\frac{\langle Q_{{\rm H}_2{O}}\rangle}{10^{28}}\,{\rm molecule~s}^{-1}}\,{\rm kg~s}^{-1},
\end{equation}
where $\langle Q_{{\rm H}_2{O}}\rangle$ is the mass production rate averaged over the orbit.
For the Jupiter-family comets studied by \citet{ahearn}, the peak
H$_2$O production rate (obtained by extrapolating observed production rates
to perihelion using the estimated dependence of $Q$ on $r$) has a median 
of $Q_{{\rm H}_2{O}}({\rm peak}) \sim 0.5\times 10^{28}$ and a range from 
(0.02--5)$\times 10^{28}$ molecule~s$^{-1}$.
To estimate the average H$_2$O production rate over the orbit, we assume
the production rate scales as $Q\propto r^{-\alpha}$, with $\alpha$ taken
from Table~V of \citet{ahearn} or set to the nominal value of 2.7; we
also set the ice production to zero when $r>3$ AU. 
Table~\ref{icetab} shows the results for comets studied both by \citet{ahearn}
and the present work.
The values of $\alpha$
are highly uncertain, because comets generally exhibit temporal variations
that are very different from power-laws. Further, the values of $Q_{\rm H_2O}$
are rather uncertain and may be statistically biased toward higher values by 
selection effects. The median $dM_{ice}/dt=9.1$ kg~s$^{-1}$ with a range
from 1.2--390 kg~s$^{-1}$. 

\begin{table*}[bh]
  \caption{ Mass-loss rates from visible observations\label{icetab}.}
  \begin{tabular*}{\tblwidth}{@{} lcccc@{} }
   \toprule
{Comet}       &  {$Q_{max}$}$^a$   &{$<Q>/Q_{max}$} &  
	{$dM_{ice}/dt$}  & {$dM/dt$(K\&K)} \\
            & ($10^{28}$ mol~s$^{-1}$) &            &  (kg~s$^{-1}$)  & (kg~s$^{-1}$) \\
\midrule
2P/Encke                    & 4.6          & 0.014      & 19     &   71       \\ 
9P/Tempel 1              & 1.7          & 0.20       & 96       &   2.7      \\ 
10P/Tempel 2            & 0.2          & 0.14       & 9.1      &   6.2	  \\ 
49P/Arend-Rigaux    & 0.18       & 0.074      & 3.8     &    0.8	  \\ 
62P/Tsuchinshan 1  & 0.26        & 0.12       & 9.4      &    0.4      \\ 
67P/Chury-Ger          & 0.41        & 0.057      & 6.7     &    1.1	  \\ 
65P/Gunn                   & 0.3          & 0.25       & 22	       &    5.6	  \\ 
69P/Taylor                 & 0.14        & 0.19       & 7.7      &    1.8	  \\ 
78P/Gehrels 2           & 0.025     & 0.17       & 1.2	       &    1.6	  \\ 
88P/Howell                & 0.24       & 0.14       & 9.5	       &    1.8  \\ 
94P/Russell 4            & 0.11       & 0.23       & 7.5	       &    3.2      \\ 
   \bottomrule
  \end{tabular*}
  $^a$ {from A'Hearn et al. (1995), Table III last column; except for
65P/Gunn for which we scaled Q(OH) from its observed distance to perihelion
 $d\theta/d\phi$ is the change in mean anomaly per unit degree on the sky.}
\end{table*}

For comparison, \citet{kk87} calculated $dM/dt$ by averaging over orbits
in a similar manner. Instead of using the ice production rate $Q_{\rm H_2O}$, they use
a rough calibration of cometary total visual magnitudes, using 1P/Halley
as a calibration source and assuming a production rate $\propto r^{-4}$.
Each revolution is calculated with its appropriate osculating orbital elements,
over the 100 yr period from 1885--1985. 
The visible magnitudes
are a measure of the total scattering by dust, and \citet{kk87} convert them
to mass loss rates assuming a gas-to-dust mass ratio of 3:1.
Their ice mass loss rates are included for the relevant comets in Table~\ref{icetab}.
The mass-loss rates inferred from the visual magnitudes and the H$_2$O 
production rates do not agree very well. Only a small part of the difference
is due to assumptions about $\alpha$; we found less than a factor of 2 variation
in $dM/dt$ for a wide range of $\alpha$, with the median changing from 
9.1 to 7.2 kg~s$^{-1}$if we set $\alpha=4$ for all comets.  

Given the wide range of $dM_{ice}/dt$ estimates, debris-to-ice ratios for
individual comets cannot be considered accurate. 
But there are enough comets in the sample that outliers
can be identified and a statistical result can be obtained. Taking the ratio
of the median $dM/dt$ and $dM_{ice}/dt$, the debris-to-ice mass ratio is 0.5.
But this ratio of medians may not be particularly meaningful,
 due to the wide range of actual
mass-loss rates for comets and the discrepancies between methods for estimating
the ice mass loss rates. There may be a wide diversity of meteoroid:ice fractions
among comets, and detailed studies of individual comets are needed to make
reliable measurements. The infrared survey results presented in this paper 
demonstrate that
meteoroids are at least comparable to ice mass production in a large sample
of Jupiter-family comets.

\subsection{Trails and `neck-lines'}

\def\degr{$^\circ$}

Brightness enhancements near a comet's projected orbit plane are not necessarily
old debris trails. The particles' orbits are only slightly different from that of
the nucleus: the radiation pressure and the in-orbit-plane components of the ejection
velocity will tend to spread particles only within the orbit plane, and only the
out-of-plane component of the ejection velocity (and gravitational perturbations
by the planets)  will cause perpendicular spread. 
Particles released with a non-zero velocity perpendicular to the comet orbit 
will still cross the orbital plane every 180\degr{} of true anomaly.  If the 
nucleus is observed at a true anomaly of $f_0$ then particles emitted at 
true anomaly $f_0 - 180^\circ$ will form a thin ridge of high column 
density when viewed at a low orbital plane angle.  The thin ridge has 
been named a 
`neck-line structure' \citep{kimura,pansecchi,fulle87}. 
A neck-line requires particles produced at true anomaly $f_0-180^\circ$, and a low
angle of the line of sight with respect to the orbit plane. 

The infrared emission in some of our comet images occupies narrow, linear
structures that are not precisely aligned with the projected orbital plane
at the time of observation. These were prime candidates for neck-line structures.
In each case, a Monte Carlo simulation was performed and the particles produced
near a true anomaly $180^\circ$ from the nucleus (at the time of observation) 
were isolated. Separate sky-plane images were made for the neck-line particles
and the remainder (debris trail particles). The debris trails follow the orbit
of the comet precisely, but the neck-lines are often at a nonzero position
angle with respect to the orbit. Experiment showed that the neck-line position angle 
was usually very close  to that of the $\beta=10^{-3}$ syndyne. 
For 67P, \citet{kelleychury}  have cleanly separated neck-line
and debris trail using infrared images, while the optical image was
dominated by the neck-line.
For
comets whose infrared emission follows the nucleus but is closer to the
$\beta=10^{-3}$ syndyne than the trail, we found the simulations are consistent
with a neck-line structure rather than strictly a debris trail.
There will
inevitably be a mixture of both the nodally enhanced neck-line and the
debris trail, from particles produced over the remainder of the orbit not
close to $f_0-180^\circ$. 
A neck-line structure
is essentially a portion of the debris trail with brightness enhanced due to
projection effects.  
The neck-lines observed in our survey have ages 
that range from 1.1 to 3.3 yr, and are observed within a few 
arcmin of the nucleus.  Therefore, these structures are 
necessarily comprised of large ($\beta < 10^{-3}$) particles.

Both neck-lines and debris trails require large particles, and the mass estimates
per unit path length $dM/d\phi$ and total mass taking into account the extent
on the sky are calculated in precisely the same manner as in the previous section. 
The neck-lines we observe are old ($>400$ days, but $<1$ orbit) and 
are composed of large particles that trail the comet, with
distinct orbits reflective of their ejection at a selected time
when their orbit planes were along the line of sight.
In theory, neck-lines are composed of particles emitted from a single 
instant, and a plot of the neck-line intensity versus distance from the 
nucleus is essentially a plot of intensity versus $\beta$.
But in practice the ejection velocities and subsequent trajectories
must be modeled relatively accurately to allow such an inversion to
the particle size distribution. Such work is beyond the scope of this
paper.
(Note also that the mass-loss rate, $dM/dt$, from the previous section
does not strictly apply to neck-lines.)
The presence of neck-lines 
for individual comets is discussed below.

\section{Descriptions of individual comets\label{sec:individual}}

{\it 2P/Encke} (Fig.~\ref{comdust1}) shows a debris trail that spans the image, as well as
a highly-elongated pseudo-coma. The {\it Spitzer} MIPS and IRAC images
are discussed in detail in \citet{reachEnckeSpitzer}; earlier results
from {\it ISO} are discussed in \citet{reachEncke}; and the
first detection from {\it IRAS} is discussed in \citet{SykesWalker}.
The brightness and inferred mass per unit mean anomaly are
relatively flat across the image. The {\it Spitzer} image measures
only a small fraction of the total mass of the trail, which
was seen to extend over 93$^\circ$ of mean anomaly in the {\it IRAS} data.
The total trail mass, estimated using the {\it IRAS}-observed mean
anomaly range and 1/2 the {\it Spitzer}-observed $dM_3/d\theta$ 
(to account for a falloff of trail brightness with distance behind
the nucleus), is $7\times 10^{13}$ g.

{\it 4P/Faye} (Fig.~\ref{comdust7}) shows a distinct debris trail leading the nucleus.
Following the nucleus, there are both a bright tail {\it and} a distinct trail,
as can be seen from the projected syndynes. 
The trail of this comet was the first to be detected optically,
extending over $10^{\circ}$ with $2^{\prime}$ width in 1991
Spacewatch images \citep{fayeIAU}. Figure~\ref{comdust7}shows the trail both leading
the and following comet, as was seen for 2P and 10P, both of 
which had long trails in the {\it IRAS} data, supporting a
common origin for the Spacewatch and {\it Spitzer} trails.

{\it 9P/Tempel 1} (Fig.~\ref{comdust1}) shows a faint debris trail due to its relatively
large heliocentric distance ($r=3.75$ AU). It was included in the survey despite
being somewhat outside our $r<3.5$ AU selection criteria, as part of preparation
for the {\it Deep Impact} mission \citep{ahearnDeepImpact}. The debris trail
was known from {\it IRAS} in 1997, and the optical depth inferred
from our data is consistent with that inferred from {\it IRAS}
\citep{SykesWalker}. Multiple-epoch observations with {\it Spitzer}
are under way to monitor the evolution of the debris trail.
The trail extends beyond the edge of the image; using the {\it IRAS}-observed
length in mean anomaly (7$^\circ$) the total mass is $1\times 10^{13}$ g.

{\it 10P/Tempel 2} (Fig.~\ref{comdust1}) shows a long debris trail similar to 2P/Encke, spanning
the image and with significant brightness leading the nucleus.
This trail was seen by {\it IRAS} in 1983, with comparable optical
depth, spanning $60^\circ$ mean anomaly  \citep{sykes90} making it
the longest and best-detected of the {\it IRAS} debris trails \citep{SykesWalker}.
Less material is leading the nucleus than following it, and the mass profile
increases somewhat toward the `following' edge of the image. 
To calculate the mass of debris leading the nucleus,
we use a constant $dM_3/d\theta=8\times 10^{11}$ g/deg 
(from the {\it Spitzer} image) and a length of $5.4\circ$ mean anomaly
(from the {\it IRAS} data).
To calculate the mass of debris following the nucleus,
we use a constant $dM_3/d\theta=1.5\times 10^{12}$
(from the {\it Spitzer} image) and a length of $60\circ$ mean anomaly
(from the {\it IRAS} data). The total mass is $9\times 10^{13}$ g.

{\it 32P/Comas Sola} (Fig.~\ref{comdust1}) presented a very large and bright coma and tail from small
particles. The coma is fan-shaped, with the opening angle not in the anti-solar
direction but rather approximately perpendicular to the projected orbit. 
The {\it dust tail} is roughly bisected by the $\beta=10^{-2}$ syndyne and bounded by 
the $\beta=10^{-3}$ and $10^{-1.5}$ syndynes. 
The {\it debris trail} can be barely discerned but appears as a 
thin linear feature, closely following the comet's projected orbit,
with a gap in azimuth between the trail and tail and in radial distance 
between the trail and coma. This phenomenon is also seen in 2P/Encke
\citep{reachEnckeSpitzer} and is due to the large particles from the
{\it present} orbit still being close to the nucleus, while the large
particles from the {\it previous} orbits are already located 
along the comet's orbit. The total extent of the debris trail is unknown,
so we estimate only the lower limit of its mass using constant
$dM_3/d\theta=3\times 10^{10}$ g/deg over a nominal 1 deg of mean anomaly.

{\it 36P/Whipple} (Fig.~\ref{comdust1}) had a small round coma centered on the nucleus and a 
$5'$ long linear feature, $15''$ wide, extending behind the nucleus. The
linear feature does not precisely follow the comet's projected orbit, as
expected for very large particles. Instead, it follows the zero-velocity 
syndyne for $\beta\simeq 10^{-2.7}$, which corresponds to particle size
$\sim 0.3$ mm (assuming density $\sim 1$ g~cm$^{-3}$), 
which we will refer to as `intermediate-sized' particles;
such particles are still within the range of radar meteor studies
and are apparently common in cometary orbits \citep{brownmeteorref}.
The position angle of the infrared feature matches that of a `neck-line'
structure due to particles produced 3.3 yr prior to observation.
The mass profile following the nucleus is fairly flat, and the total extent
of the emission is unknown, so we estimate a lower limit to the mass assuming a 
nominal $1^\circ$ length in mean anomaly and correcting to the slightly smaller
particle size
$M > dM_3/d\theta \times 10^{2.7-3} = 3\times 10^{11}$ g.

{\it 42P/Neujmin 3} (Fig.~\ref{comdust2}) had a faint tail, $30''$ wide, extension following 
the nucleus. This is apparently a dust tail, as it does not follow
the projected orbit. Enough area was covered, and the anti-solar direction
was nearly opposite the projected orbit direction, so that an upper limit
$I_\nu<0.02$ MJy~sr$^{-1}$ 
could be placed on the debris trail using the portion of the image
away from the coma and tail.

{\it 48P/Johnson} (Fig.~\ref{comdust2}) was bright and had one of the best viewing geometries, 
and cleanest separation between trail and tail, in the survey. 
The fan-shaped dust coma is roughly bounded by the $\beta=10^{-1}$
to $10^{-3}$ syndynes.
The debris trail is clearly distinct from the coma and follows
the projected orbit very precisely, so it can only be composed of
large particles with $\beta<10^{-4}$. 
The trail is $17''$ wide; such narrow trails, frequently found
in this survey, would have been highly beam diluted to {\it IRAS}
so are not surprisingly not detected. The trail is fainter,
but present, leading the comet, requiring non-zero ejection velocity.
The trail extends to at least the edge of the image; an estimated
lower limit to its mass, assuming constant mass per unit mean anomaly
and a length of $1^\circ$ mean anomaly,
is $2\times 10^{11}$ g.

{\it 49P/Arend-Rigaux} (Fig.~\ref{comdust2}) has a bright point-like nucleus with diffraction
rings and a dust tail due to small particles, $\beta=10^{-1}$ to $10^{-2}$.
There is no debris trail, despite excellent viewing geometry and high-quality
data. We place an upper limit on the surface brightness, $I_{\nu}<0.03$ MJy~sr$^{-1}$ 
corresponding to an optical depth $\tau<2\times 10^{-10}$, which is lower than
that of detected trails.
Because the comet was close, the trail is expected to be wider than those
for most other comets ($\sim 1'$, scaling other comets trail widths by $\Delta$),
but a very wide area ($30'$) was mapped and should have included the debris
trail. The upper limit on the debris trail is therefore very strict, and
an explanation for the lack of trail is required. As discussed below, comets with
more non-asteroidal orbits (lower Tisserand parameter), 
like 49P, seem to have weaker trails. 
It is quite possible that the close approach to Jupiter in 1997 is
the root cause for the lack of a detectable trail.
Particles from the previous revolution are normally the ones that
would be detected in the debris trail on this revolution.
The 1997 close approach may have perturbed the orbits of large particles from 
the previous revolution, 
scattering them to a different location or dispersing them widely.

{\it 53P/van Biesbroeck} (Fig.~\ref{comdust2}) had a poor viewing geometry, with the tail and
trail superposed to within $3^\circ$ position angle. There was no debris
leading the comet, down to an upper limit of 0.05 MJy~sr$^{-1}$ surface 
brightness, which corresponds to $\tau<9\times 10^{-10}$.

{\it 56P/Slaughter-Burnham} (Fig.~\ref{comdust3}) has one of the most clear-cut debris trails,
with an excellent viewing geometry. The debris trail is narrow, extends both
leading and following the nucleus, precisely following the projected orbit
as expected for particles with $\beta\ll 10^{-3}$ experiencing negligible
radiation pressure. The lower limit to the trail mass, for $\beta=10^{-3}$
particles and a length of $1^\circ$ mean anomaly is $8\times 10^{10}$ g.
        
{\it 62P/Tsuchinshan 1} (Fig.~\ref{comdust3}) was observed to have both a tail,
roughly bounded by the $\beta=10^{-3}$ to $10^{-1}$ zero-velocity 
syndynes, and a debris trail. The trail is separated from
the tail following the comet, thanks to a favorable viewing
geometry. It is however faint, and the trail mass can only be 
crudely bounded, if we assume $>1^\circ$ mean anomaly length,
as $M>5\times 10^{10}$ g.

{\it 65P/Gunn} (Fig.~\ref{comdust3}) was observed with a very bright tail following the
comet as well as a prominent debris trail leading the comet.
The dust tail is slightly shifted from the orbit, mostly consistent with
the zero-velocity syndynes of $\beta=10^{-2}$ to $10^{-3}$, but
including the projected orbit ($\beta=0$). The
debris trail leading the comet is closely aligned with the orbit.
This trail had been previously detected by {\it IRAS} 
\citep{SykesWalker}. The trail extends to the edge of the {\it Spitzer}
image without decreasing significantly in brightness. 
Using the {\it IRAS}-observed lengths of $0.3^\circ$ mean anomaly
leading and $5.9^\circ$ following the nucleus, together with
the {\it Spitzer}-observed brightnesses, yields a total mass 
$9\times 10^{12}$ g.

{\it 67P/Churyumov-Gerasimenko} (Fig.~\ref{comdust3}) had previously been 
found to have a debris trail
by {\it IRAS} \citep{SykesWalker}. The {\it Spitzer} observation is discussed
in detail in a separate paper \citep{kelleychury}. The portion of the debris 
trail leading the comet is evident, while the portion following the comet is
contaminated by a neck-line structure and is therefore not solely a debris trail.
The mass of the trail can be roughly estimated using its brightness leading
the nucleus and the {\it IRAS}-observed length, yielding $1\times 10^{12}$ g.

{\it 69P/Taylor} (Fig.~\ref{comdust4}) has a bright tail with a range of particle sizes,
with the bulk falling within the $\beta=10^{-2}$ to $10^{-3}$
zero-velocity syndynes. A debris trail precisely aligned with the
orbit is clearly evident leading the comet; it is confused with the
tail following the comet. \citet{kresak} 
discussed the case of 69P/Taylor, which was observed split in 1916 but 
was subsequently perturbed by Jupiter in 1925 so that debris from the
splitting should be widely dispersed. He mentions that a trail should 
be present but below the detection limit of {\it IRAS}. 
Our new observations
are more sensitive, allowing us to detect the debris trail with 
{\it Spitzer}. If the comet were only discovered a few years
ago, we would never know it had split in the past; this is
an example of cometary calving, which is apparently
frequent \citep{jewittsplitting}.
The total extent of the trail is not known, so we only estimate a
rough lower limit for $1^\circ$ mean anomaly of $M>2\times 10^{10}$ g.

{\it 71P/Clark} (Fig.~\ref{comdust4}) was detected as a long debris trail, but
without a bright nucleus or coma at the predicted location (using the orbit from
1998). Using the updated orbit after recovery in 2005 \citep{pittichova},
 a compact source at the nucleus' position
is indeed present $2'$ ahead of that predicted from the 1998 orbit;
this source is moving at the predicted rate based on the 2006 ephemeris. 
The most
fascinating result for this comet is that its debris trail is long and even
increases in brightness following the comet, peaking $11'$ behind it. This
cannot be explained by a simple dust production history. The $10^6$ particle
Monte Carlo simulation (with $r^{-2}$ dust production) for this comet shows only 
a coma and trail, so the enhancement in trail behind the comet is not due to
a geometric projection effect. The mass per unit mean anomaly seems to increase with
distance following the nucleus, and the trail extent is unknown. 
A lower-limit mass for $1^\circ$ mean anomaly is $8\times 10^{11}$ g.
In 2006, this comet was observed again at much closer distance; Figure~\ref{comdust7}
shows a clearly-detected trail leading the nucleus. The trail following the nucleus
cannot be cleanly discerned from a possible tail until near the edge of the image,
at which point we see that the narrow ridge of emission extending across the entire
image is indeed the debris trail, following closely the projected orbit of the comet.

{\it 78P/Gehrels 2} (Fig.~\ref{comdust3}) was exceptionally bright when observed,
with a bright tail due to particles of indeterminate size (due
to the unfavorable viewing geometry with the anti solar direction
nearly parallel to the orbit-following direction). Leading the
comet, there is a narrow feature that lies precisely along the
projected orbit, which we identify as the debris trail.
Using the mass per unit mean anomaly leading the comet, and
a lower-limit length of $1^\circ$ mean anomaly, yields a
mass $M>8\times 10^{10}$ g.

{\it 88P/Howell} (Fig.~\ref{comdust4}) was observed when it was relatively closeby,
with unfavorable geometry (tail and trail overlapping), and covering a relatively small
field of view. The extremely bright tail is likely due to small particles.
The only portion of the image free of small particles is leading the comet, where 
there is a faint linear feature parallel to but slightly shifted (by less
than its width) from the projected orbit of the comet. Based on comparison
to images of other comets (78P, 123P) viewed similarly and with clear `leading'
debris trails despite bright tails, we tentatively identify this feature 
as Howell's debris trail. The lower-limit mass, derived as for 78P, is $M>8\times 10^{10}$ g.

{\it 94P/Russell 4} (Fig.~\ref{comdust4}) has a moderately bright debris trail following the nucleus,
with no detectable emission leading the nucleus. The {\it Spitzer} observation
was adversely affected by an observation of Saturn immediately preceding,
but the image was adequately recovered. The trail is close to the
projected orbit but follows the $\beta=10^{-3}$ syndyne better. Based on
comparison to simulations, much of the infrared feature is likely to be a neck-line
structure due to particles produced 2.0 yr prior to observation.
The estimated mass of the structure $>3\times 10^{10}$ g.

{\it 103P/Hartley 2} (Fig.~\ref{comdust5}) had a bright tail, roughly bounded by the
zero-velocity syndynes for $\beta=10^{-3}$ and $10^{-1}$. 
The separation 
between the tail and trail was large enough to set an upper limit to
the trail brightness relatively far behind the comet; no trail was
seen leading the comet with an upper limit $I_{\nu}<0.1$ MJy~sr$^{-1}$.

{\it 104P/Kowal 2} (Fig.~\ref{comdust5}) had extensive infrared emission following the nucleus,
but the apparent trail does not precisely follow the comet's orbit.
Instead, it is closer to the $\beta=10^{-3}$ syndyne and may contain
significant contribution from intermediate-sized particles. The
mass estimate $M>2\times 10^{11}$ g is a lower limit because of
the unknown total extent, though the surface brightness does decrease
significantly before the edge of the image (suggesting most of the mass
was observed). The orientation of the infrared feature is similar to 
that expected for a `neck-line' from particles ejected 1.1 yr prior
to observation, but the feature is broader than a neck-line.
The feature is more likely a tail, due primarily to intermediate-sized particles.

{\it (4015) 107P/Wilson-Harrington} was not strictly part of the short-period comet
survey but was observed in the same manner. 
It was found to be a point source, with no extended emission
along the orbit to a limit of $I_\nu < 0.05$ MJy~sr$^{-1}$.
At the observed heliocentric distance $r=2.25$ AU, this corresponds
to an optical depth limit $\tau<4\times 10^{-10}$.

{\it 108P/Ciffreo} (Fig.~\ref{comdust7}) shows a narrow trail in the orbit of the comet.
The bright source at the predicted location of the nucleus looks point-like and may
actually be a celestial source coincidentally close to the ephemeris position. A
smaller, somewhat resolved source is located following the predicted location of
the nucleus and within the debris trail. This object appears to be the actual nucleus,
at least in terms of debris trail production, based on the location of some debris leading
and most following its location. While the syndynes in Fig.~\ref{comdust7}
are centered on the predicted location of the nucleus, the flux
reported in Table~\ref{trailtab} is that of the smaller source.
More detailed modeling is required to 
address these issues.

{\it 111P/Helin-Roman-Crockett} (Fig.~\ref{comdust5}) has a thin trail extending to the edge 
of the image and precisely following the comet's orbit. The brightness
does not decrease significantly at the edge of the image, suggesting
the trail may be massive; its narrow width shows it is also devoid
of small particles. An estimate lower-limit to the mass 
$M>3\times 10^{11}$ g.

{\it 116P/Wild 4} (Fig.~\ref{comdust5}) was observed when relatively
closeby and presented a very bright dust tail that makes it
impossible to separately locate a possible debris trail.

{\it 120P/Mueller 1} (Fig.~\ref{comdust5}) has a prominent debris trail both
leading and following the nucleus, closely following the orbit. 
Assuming a length at least $0.3^\circ$ leading
and $1^\circ$ following the nucleus, the trail mass 
$M>2\times 10^{11}$ g.

{\it 121P/Shoemaker-Holt 2} (Fig.~\ref{comdust5}) had a bright elongated coma/tail due
to particles of indeterminate size.
While the anti solar and comet-following angles were not widely 
separated, a long enough field was observed to allow a debris trail 
in the comet's orbit to be distinguished from a small-particle-only tail
far from the nucleus.
There is no clear, contiguous debris trail in the comet's orbit, as
there was for most of the other comets in the sample. However, there
were two patches of emission along the comet's orbit. The first patch
is circular, located $9.8'$ following the nucleus, and the second
one is elongated along the orbit, approximately $15.8'$
following the nucleus; their brightness $\sim 0.2$ MJy~sr$^{-1}$
corresponds to an optical depth $\sim 2\times 10^{-9}$. While
these patches are aligned along the orbit, it is very likely (and
cannot be determined with the present data) that these patches are 
small interstellar clouds. Inspecting the {\it IRAS} images
as reprocessed by \citet{iris}, there is significant,
structured interstellar emission in the field.

{\it 123P/West-Hartley}  (Fig.~\ref{comdust6}) was observed relatively close by and had
a very bright tail of small particles. The debris trail following the comet can
be barely distinguished from the tail. Leading the comet, the debris trail 
precisely along the projected orbit, is detected without confusion from smaller 
particles. The presence of a significant trail leading the comet suggests the
debris trail may be massive, but from the present observations only a rough
lower limit can be made, by assuming a length $0.3^\circ$ leading
and $1^\circ$ following the nucleus; this yields $M>5\times 10^{10}$ g.

{\it 127P/Holt-Olmstead} (Fig.~\ref{comdust6}) was observed under very
favorable geometric conditions, with the small-particle production
being minimal at $r=2.74$ AU and the anti solar (tail) direction 
$169^\circ$ from the trailing direction. 
The brightness decreases significantly before the edge of the image,
suggesting we have observed the bulk of the debris (though there could always
be a fainter trail of larger particles). These factors combine to allow
an actual measurement (rather than a limit) of the mass 
of $M \simeq 4\times 10^{10}$ g. 
The extended emission stretches
approximately along the orbit, completely inconsistent with a small-particle
tail. But the orientation is not precisely along the orbit, being closer
to the $\beta=10^{-3}$ syndyne. The orientation of the infrared emission matches
very well that of a `neck-line structure' due to particles ejected
1.7 yr prior to observation.

{\it 129P/Shoemaker-Levy 3} (Fig.~\ref{comdust6}) had one of the best-defined
debris trails owing to its brightness and favorable viewing geometry. The
trail extends to the edge of the image without decreasing significantly
in brightness, so only a lower limit to the mass can be made, assuming
trail length at least $0.3^\circ$ leading and $1^\circ$ following the 
nucleus in mean anomaly, yielding $M > 5\times 10^{11}$ g.

{\it 131P/Mueller 2} (Fig.~\ref{comdust6}) had very faint extended emission,
evident only after smoothing the image. This faint emission is only present
following the comet and would correspond to a lower limit to the trail mass
assuming $1^\circ$ length in mean anomaly of $M>2\times 10^{10}$ g.

{\it 133P/Elst-Pizarro} has a very faint
($\sim 0.02$ MJy~sr$^{-1}$), narrow debris trail confined very precisely to 
the comet's orbit. This comet has an orbit similar to main belt asteroids and
exemplifies the class as defined by \citet{hsieh}. The debris trail appears
similar to other narrow trails in our sample, suggesting that the main belt
comet lose significant mass in the form of large meteoroids, just like the
Jupiter-family comets. More detailed modeling and image analysis are
under way to better constrain mass production history.

{\it P/2003 S2} (Fig.~\ref{comdust6}) had a debris trail precisely
following the comet's orbit. The brightness decreases significantly
by the edge of the image, allowing a mass estimate
$M\simeq 4\times 10^{10}$ g,
although faint trail emission extends to the edge of the image (so
the total trail mass could be significantly higher).
Part of the infrared emission may be due to a neck-line structure;
it is not possible to separate them due to the viewing
geometry placing the trail and neck-line on top of one another.

\section{Statistical Trends of debris trail properties}

We summarize properties of each comet that might influence the
properties of their debris trails in Table~\ref{trailtab}.
The orbital size and eccentricity are listed, together with
the Tisserand invariant for
gravitational scattering by Jupiter, $T_J$ \citep{tissref}.
And the date of the last passage of a comet into Jupiter's gravitational
sphere of influence is $T_{pert}$.

\begin{figure}
    \includegraphics[width=0.49\textwidth]{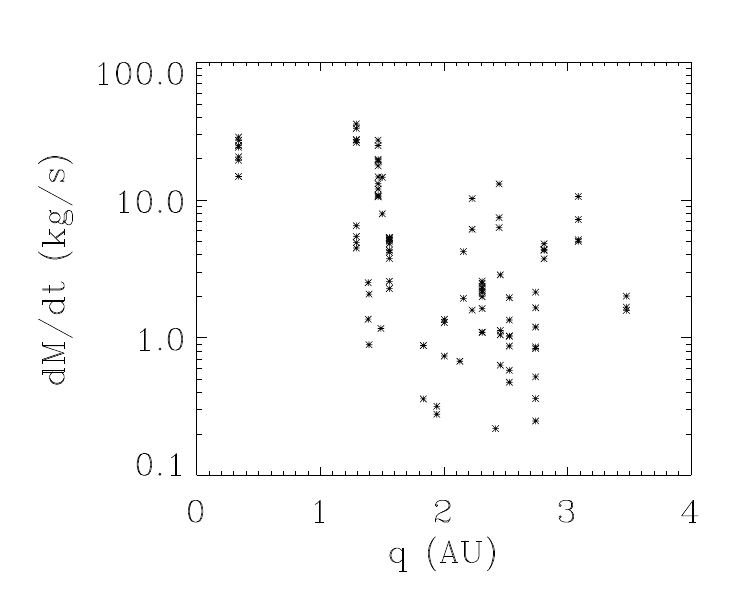}
    \caption{Scatter diagram of the mass loss rate required to produce
the observed debris trails versus perihelion distance.
\label{dmdtplot}}
\end{figure}

The amount of material in a given comet's debris trail may be correlated with
the present shape of the comet's orbit. 
Figure~\ref{dmdtplot} shows the debris mass production rate versus $q$. 
Weak correlations are present
between $\dot{M}$ and $a$ or $q$. The correlation coefficients 
between $\dot{M}$ and $a$ or $q$ are -0.64 and -0.32, respectively;
while the correlation coefficients between $\log \dot{M}$ and $a$ or $q$
are -0.66 and -0.62, respectively.
This apparent correlation is unlikely to be purely a selection effect.
While comets observed at large $r$ could only be detected if they have
relatively larger $\dot{M}$ (at fixed sensitivity limit), our survey does
not appear to be sensitivity limited for trail detection. Indeed the comets
with no detected trails tended to be observed at smaller $r$.
Further, the comets with small $q$ were not preferentially observed at 
small $r$ (where they would be brighter).

The correlation with orbital shape ($q$ or $a$) can also be expressed
in terms of the Tisserand invariant for Jovian perturbations, $T_J$. 
Comets with little interaction with Jupiter
(asteroidal or Encke-type orbits, $T_J> 2.94$) have a median
$dM/dt=3.4$ kg~s$^{-1}$, while those with $T_J<2.94$ have 
a median $dM/dt=0.8$ kg~s$^{-1}$.
(A clear outlier is 67P, which is the only low-$T_J$ comet with
$dM/dt > 2$ kg~s$^{-1}$.)
These trends suggest that comets spending more
time closer to the Sun may produce more debris, and
comets dynamically decoupled from Jupiter may retain debris
longer.

The general trend of more-massive trails for
comets with smaller $q$ 
goes somewhat counter to what might be expected from the
trend of dust-to-gas ratio being {\it lower} at smaller $q$ as
found in the visible-light survey by \citet{ahearn}. 
However, if the grains being detected in the inner coma photometry,
traced by the scalar $Af\rho$ by \citet{ahearn} are predominantly
small grains, perhaps from fresher surface material, then the
new result on debris trails could be explained as an evolution of the
{\it size} of the material ejected from comets. Comets spending
more time in the inner Solar System may be dynamically older and 
more depleted in small particles,
while they still appear capable of ejecting larger particles,
which build up in their orbits.

For the observations presented in this paper, there is no clear correlation 
between the amount of debris in an orbit and the date of last perturbation.
Such a correlation would be expected if the perturbations brought comets
that were previously in outer-solar-system orbits into $q<2$ AU orbits,
because orbits with $q>2$ might be expected to have 
less devolatilized crust. 
The date of the last significant perturbation was after 1950
for 7 comets: 62P, 65P, 67P,
103P, 111P, 116P, and 120P. 
Of these, one has a clear non-detection of the trail.
It was between 1900 and 1950 for 6 comets
(32P, 36P, 69P, 78P, 88P, and 129P).
Of these, one has a clear non-detection of the trail.
For 17 comets, the strong perturbations occurred before 1900.
For the 11 of these relatively stable orbits with detected trails,
the average mass production rate is 5.9 kg/s. One of the
17 stable orbits has a clear non-detection of the trail. 
We also checked whether weaker perturbations due to passing within
1 AU of Jupiter have a significant effect on the amount of material
in the near-nuclear trails.
Such perturbations are generally minor; they are 
enough to shift material produced before the perturbation discernibly 
away from the current projected orbit but not enough to disperse the
particles into the zodiacal cloud. (The change in $q$ was typically of
order 0.1 AU.) 
Thus the date of the last strong perturbation does not have a
dramatic influence on the amount of debris trail material
close to the nucleus, as traced by our {\it Spitzer} survey. 
We do expect a significant correlation between $T_{pert}$ 
and the trail {\it length}; however, the present observations do not cover
enough of each comet's orbit to show the ends of the trails.

\begin{figure}
    \includegraphics[width=0.49\textwidth]{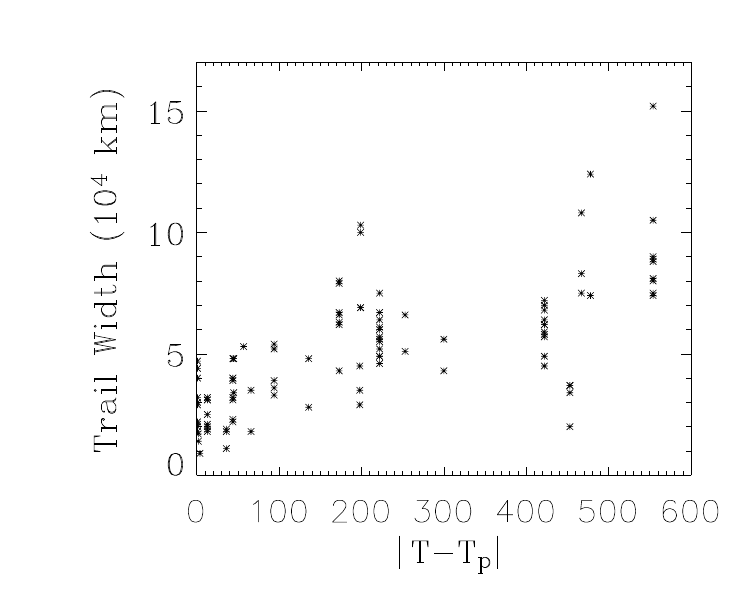}
    \caption{Correlation of the debris trail width versus
the time of observation relative to perihelion (days).
\label{widthplot}}
\end{figure}

The widths of the debris trails also show some systematic trends.
Figure~\ref{widthplot} shows the trail width versus the time since
perihelion.
The trails observed furthest from perihelion (i.e. larger $|T-T_p|$ or $r$)
are wider. 
If the observations were sensitive to particles emitted within months
of the time of observation, the observed trend would be counter
to expectations: particles released
further from perihelion would be moving slower and would remain
closer to the midplane.
However if the particles are years old, which we believe to be the
case, and they are mostly produced near perihelion, then 
the width should increase with time since perihelion:
\begin{equation}
W \propto V_{\perp} |T-T_{p}|,
\end{equation}
where $V_{\perp}$ is the component of
the ejection velocity perpendicular to the comets' orbits.
Gravitational perturbations would further increase the trail width for older particles, but on timescales of multiple orbits.
If we interpret the increasing width as due only to ejection
velocity, the slope of the trend in Figure~\ref{widthplot} yields 
$V_{\perp}\sim 2$ m~s$^{-1}$.
Using equation~\ref{equation:vej}, assuming particles are mostly produced near perihelion, with angle $\psi$ relative to the orbit
plane, we estimate the particle size
\begin{equation}
\beta \sim \left(\frac{v_{\perp}\cos\psi}{v_{1}}\right)^{2} q(AU),
\end{equation}
yielding $\beta\sim 5\times 10^{-6}$ for a typical perihelion 
$q\sim 1.5$ AU, ejection angle $\psi=30^{\circ}$, and
$v_{1}=1$ km~s$^{-1}$.
This size estimate is in accord with the results presented
elsewhere in this paper, which show $\beta\ll 10^{-3}$. 
For the trail width trend to be produced by particles with
$\beta$ as large as $10^{-3}$ requires $v_{1}=0.07$ km~s$^{-1}$,
a value too slow to explain the morphology of 2P/Encke's coma
\citep{reachEncke} and too low to explain the lengths of
trails seen by {\it IRAS} \citep{SykesWalker}.

\section{Conclusions}

The {\it Spitzer}/MIPS images presented in this paper 
demonstrate that the production of mm-sized debris is
a common feature of Jupiter-family comets: the orbits of
most (at least \ncomtrail\ out of the \ncomobs\ in the present survey)
comets are delineated on the sky by `trails' of
mid-infrared emission. 
The debris trails can only be due to particles
with a small ratio of radiation pressure to gravity, 
$\beta< 10^{-3}$. 

The extended surface brightness near Jupiter-family comets
contains a mix of particle sizes, $1<\beta<10^{-3}$
($\sim1$--1000 $\mu$m for density $\sim 1$ g~cm$^{-3}$).
Generally, the comae are much brighter in the regions traveled 
by large particles, bounded by the $10^{-2}<\beta<10^{-3}$
(100-1000 $\mu$m) syndynes,
than in the regions traveled by smaller particles. 
Examples of observations of large-particle-dominated comae 
include 2P/Encke and 48P/Johnson, for which the viewing geometry
allowed straightforward separation of particles of different size. 

Dust {\it tails}, stretching roughly between the anti-solar direction 
and the $\beta=1$ syndyne, were detected from some comets,
indicating that small particles can also be detected in
the mid-infrared images when they are present. Tails are
routinely detected for comets observed closest to the Sun.
The strong heliocentric and comet-to-comet
variety of tails, in contrast to the prevalence 
and similarity of trails suggests a variation in the 
dust size distributions---specifically, the ratio of
surface area in large versus small particles---among comets.
Further work is required to determine whether it is the
actual surface area of grains that is varying or the manner
in which they are produced (i.e. the velocity distribution
or efficiency of fragmentation).

Debris trails are massive, and the inferred mass production
rate of mm-sized debris is larger than that of sublimating ice.
The predominance of large particles in cometary mass loss 
agrees with {\it in situ} spacecraft observations of cometary particles.
\citet{greenWild2} showed that the mass distribution detected during 
all three the comet encounters with dust monitors is completely
dominated by large particles. We separate
the particles into three populations based on the shape
of the mass distribution: 
{\it small particles},
smaller than $10^{-6}$ g (size less than 50 $\mu$m, $\beta>10^{-2}$), 
have a cumulative mass distribution
$N(<m)\propto m^{-0.75}$; {\it intermediate particles}
above this mass and until $\sim 10^{-4}$ g 
(size 250 $\mu$m, $\beta\sim 10^{-3}$)
have relatively constant cumulative fluence; 
then for {\it large particles} the cumulative mass distribution 
decreases again, approximately as $m^{-0.75}$.
Such a size distribution is consistent with the {\it in situ} 
observations during the {\it Giotto}
 encounter with 1P/Halley \citep{mcdonnellHalley}
the {\it Stardust} encounter with 81P/Wild 2 \citep{greenWild2},
and interplanetary meteoroids detected by near-Earth spacecraft
and meteor magnitudes \citep{grun85}.
The mass is contained in the largest particles for this size distribution
(both overall and within each of the three populations).
Within the population of small or large particles, the surface area 
is contained in the smallest particles. But for the overall distribution,
the surface area is dominated by the smallest of the `large' particles,
i.e. those with  $m\sim 10^{-4}$ g, size $\sim 250$ $\mu$m, 
and $\beta\sim 10^{-3}$. 

\begin{figure}
    \includegraphics[width=0.49\textwidth]{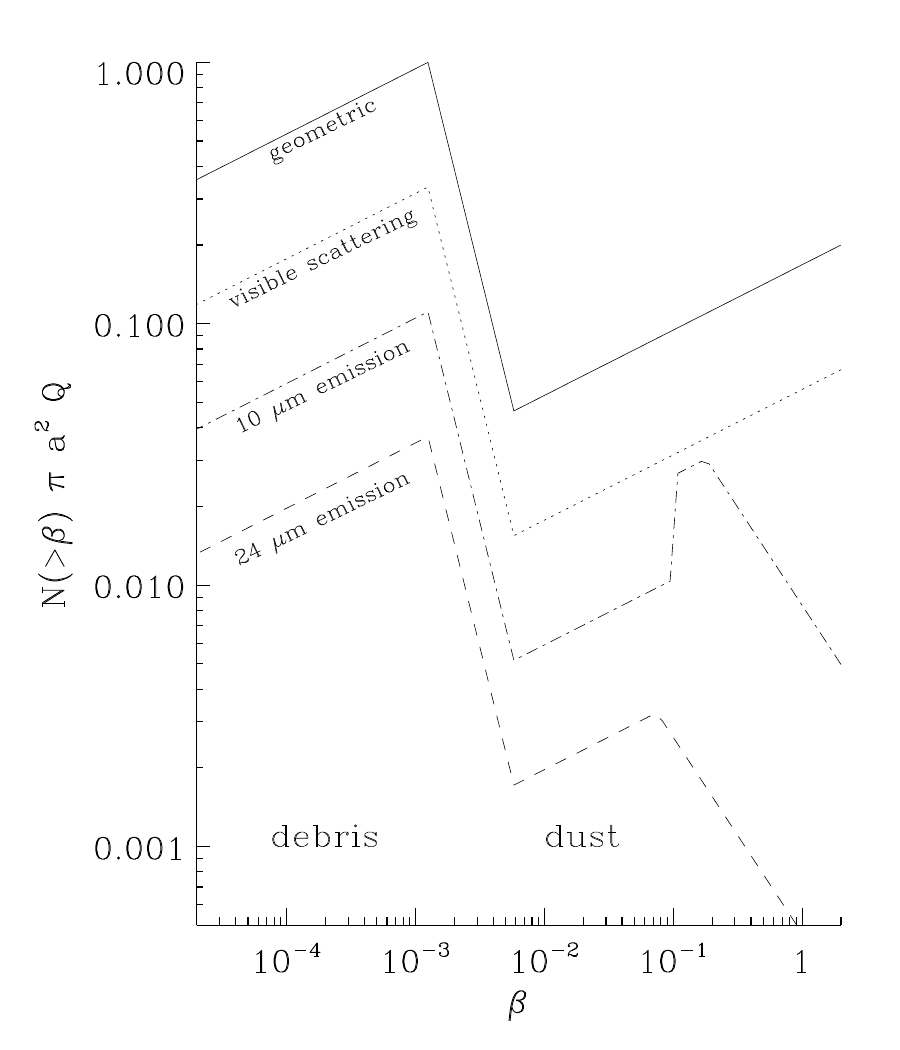}
\caption{Contribution of different sizes
to the surface area of cometary 
particles, for a three-population size distribution similar
to that found from the {\it in situ} observations by
dust detectors on the {\it Stardust} spacecraft in
the coma of 81P/Wild 2. 
Particle sizes on the horizontal axis are parameterized by
$\beta/Q_{pr}\rho=0.57 \mu{\rm m}/s$, where $s$ is the particle
radius, $Q_{pr}$ is the radiation pressure efficiency in the
solar radiation field ($\sim 1$ for particles with $\beta<0.1$)
and $\rho$ is the particle density in g~cm$^{-3}$.
The solid line shows the 
fractional contribution to geometric area; the dotted line
shows the contribution to visible light scattering; the
dashed line shows the contribution to 24 $\mu$m emission.
The dash-dot line shows the Planck-weighted 
contribution to 10 $\mu$m emission
(such as the silicate feature), if the smaller grains 
($\beta>0.1$) have higher temperature than the larger grains
(340 K {\it vs} 280 K).
All curves overlap for small $\beta$
(i.e. on the left side of the plot); then, one by one, each
effective cross section diverges from the geometric cross-section,
with 24 $\mu$m emission diverging at lowest $\beta$ and visible
scattering diverging at highest $\beta$. 
\label{sizedist}}
\end{figure}

Figure~\ref{sizedist} shows the size distribution weighted by the
effective surface area.
For 24 $\mu$m emission, the effective surface area is
the cross-sectional area times the absorption efficiency
$Q_{24} = (2 \pi a / \lambda)$ for $a < \lambda/2\pi$
with $\lambda=24$ $\mu$m.
For scattering, $Q = (2 \pi a / \lambda)^4$
for $a < \lambda/2\pi$ with $\lambda=0.55$ $\mu$m for visible
light and 1 cm for radar. The efficiencies are all set to unity
for $a>\lambda/2\pi$.
Using the size distribution observed 
within the coma of 81P/Wild 2 \citep{greenWild2}, the effective surface area for mid-infrared
emission is strongly dominated by the particles of $\beta\sim 10^{-3}$,
while the effective surface area for visible-light scattering has
a significant but not dominant contribution from smaller particles.
(Radar backscatter would be produced only by the large particles.)
The nature of the size distribution, which is {\it not a single power law
over the range of sizes from micron to cm},
explains why the debris trail observations are dominated by
particles with $\beta>10^{-3}$. 

There is an apparent conflict between our interpretation of
the mid-infrared images (and the {\it in situ} size distributions)
and the traditional interpretation of optical images and infrared spectra.
The optical imaging and infrared spectroscopy 
studies generally derive particle sizes of order
10 $\mu$m and consider particles only up to 100 $\mu$m
\citep[cf.][]{kelley06,lisseencke,kolokolova}, which all fall
within the {\it small particle} population.
The existence of some small particles is required to explain
silicate emission features ($a < \lambda/2\pi$ where the
silicate feature is at $\lambda\sim 10$ $\mu$m).
Silicate features have also been
correlated with the `super-heating' of grains above the isothermal
sphere (blackbody) temperature, due to the lower absorption
efficiency of small grains in the infrared where their cooling
occurs ($a < 65 / T$ $\mu$m where the grain temperature $T$ is in K).
Silicate features are generally not strong for Jupiter-family
comets, and they can arise from a population of grains distinct from the
large particles that dominate the mass loss and produce the debris trails.
Figure~\ref{sizedist} shows that a significant contribution to
mid-infrared emission (and in particular the silicate feature)
can arise from the the small particle population if they are 
significantly hotter than the large grains,
even if a large particle population (in excess of the 
power-law fitted to the small particles) is included. 
Therefore, we suspect that previous models for coma spectra and
optical scattering are sampling only the small particle
population; the abundance of large particles 
such as in meteoroid streams is far larger
than would be inferred from an extrapolation of 
a simple power law normalized to the small particle abundance.


That large particles are produced by most comets is now beyond doubt, but
the total mass and the distribution of large particles over comets' orbits
remains largely unknown. The present survey only covers the near-nuclear environment,
with the debris trails dominated by material from this and the previous 
revolution. The slow spread of debris trail material, and the rapid
influence of gravitational perturbations, spreads debris over a wide
area. 
Strong perturbations by Jupiter occur every $\sim 10^2$ 
yr for Jupiter-family comets.
Comets in the most stable orbits can build up a meteoroid
complex, containing many orbits' debris, easily recognizable from surface brightness imaging, while comets suffering recent 
perturbations will have only young particles in their trails.
The debris detected in the present survey with {\it Spitzer} was
faint for most comets, $<1$\% of the zodiacal light surface brightness
and with a width $<50''$. Such features are exceptionally difficult
to detect in visible light \citep[e.g.][]{ishiguro}
and shallow, wide-area surveys (e.g. {\it IRAS}).
Deeper investigations along the orbits of comets may reveal
the extent of their meteoroid complexes and the total mass
production of large particles.

The meteoroids of some comets enter the Earth's atmosphere as
meteor showers \citep{jenniskens}. Meteor showers are much more widely
dispersed than the debris trails we observed. The debris trails
are actually more comparable to the meteor {\it storms} 
\citep{jenniskens,kresak}. 
The debris trails evolve into the wider
meteoroid streams and gradually into the sporadic meteors.
The mass of meteoroid streams has been estimated from observed
meteor rates, yielding $10^{15-17}$ g for four well-studied
streams \citep{hughesmcbride}. These masses are much larger 
than those
we infer from most debris trails,
with the lowest estimated meteoroid stream mass still an order of
magnitude larger than the largest estimated debris trail mass.
The well-known meteor streams are likely to arise from meteoroids
produced by the largest and longest-lived comets that cross
the Earth's orbit. Their properties provide a relatively unique,
long-term view of cometary mass loss that is complementary
 to that obtained from studying debris trails.

The meteoroid mass production rates measured from the present
survey has a median 2 kg~s$^{-1}$ per comet,
with a few undetectable ones having smaller mass-loss.
There are $\sim 200$ known active comets with orbits in the Jupiter family.
Assuming the comets we observed are a representative
sample, and allowing for as many as 20\% having no current meteoroid
production (based on the non-detection rate),
the total meteoroid input from short-period comets is
$\sim 300$ kg~s$^{-1}$. The debris trail particles are all
on bound orbits,
very similar to those of their parent comets---though they will
gradually scatter
away from their parent comets due to their slightly different initial
orbit, non-zero $beta$,
different non-gravitational forces due to outgassing and Yarkovsky effect,
and different gravitational perturbations by the planets. Thus debris trail
particles will gradually fill the inner Solar System. Their lifetime
is mostly limited
by mutual collisions with other interplanetary particles
\citep{grun85}; the resulting
collision products would form via collisional cascade a size
distribution including
the $\sim 20$--200 $\mu$m sized particles that generate the zodiacal light
\citep{grun85,reach88}.

\citet{leinert83} estimated that the zodiacal cloud loses 
600-1000 kg~s$^{-1}$ due to
mutual collisions and Poynting-Robertson drag, requiring a
corresponding continuous
input to maintain the zodiacal light at constant brightness. 
Constancy of the zodiacal light is not actually required, with the best
observational evidence
being constancy to $\pm 2$\% over 11 yr of Helios observations
\citep{leinert89}.
A cometary contribution to the zodiacal cloud may be highly variable,
in particular if it were dominated by a few very massive comets 
soon after injection into
a low-perihelion orbit \citep{napier01}, or it
may be relatively constant if the mass input is relatively evenly distributed
among the population of short-period comets as part of their gradual
disintegration.
It appears, based on the available evidence, that short-period comets
lose enough mass to contribute significantly to the interplanetary dust 
complex. Long-period comets and asteroids
almost certainly provide additional contributions.

\section*{Acknowledgments}

WTR thanks Giovanni Fazio, who as IRAC Principal Investigator supported
part of this project through his Guaranteed Time allotment.
This work is based on observations made with the Spitzer Space
Telescope, which is operated by the Jet Propulsion Laboratory, California
Institute of Technology under a contract with NASA. Support for this work was
provided by NASA through an award issued by JPL/Caltech.

 

\end{document}